\documentclass[a4paper,11pt]{article}
\usepackage{jheppub} 
\usepackage{cancel}
\usepackage{comment}

\title{\boldmath Iyer-Wald ambiguities and gauge covariance of Entropy current in Higher derivative theories of gravity}

\author[a]{Alokananda Kar,}
\affiliation[a]{Department of Physics, Indian Institute of Technology (Indian School of Mines) Dhanbad, Jharkhand 826004, India}

\author[b]{Prateksh Dhivakar,}
\affiliation[b]{Department of Physics, Indian Institute of Technology Kanpur, Kalyanpur, Kanpur 208016, India}

\author[c]{Shuvayu Roy,}
\affiliation[c]{School of Physical Sciences, National Institute of Science Education and Research, \\
An OCC of Homi Bhabha National Institute, Jatni-752050, India}

\author[a]{Binata Panda,}

\author[a]{Anowar Shaikh}

\emailAdd{21dr0013@phy.iitism.ac.in,prateksh@iitk.ac.in,shuvayu.roy@niser.ac.in,\\
binata@iitism.ac.in, anowar.19dr0016@ap.iitism.ac.in}

\abstract{In \cite{Bhattacharyya:2021jhr,Biswas:2022grc} [arXiv:2105.06455, arXiv:2206.04538], the authors have been able to argue for an ultra-local version of the second law of black hole mechanics, for arbitrary diffeomorphism invariant theories of gravity non-minimally coupled to matter fields, by constructing an entropy current on the dynamical horizon with manifestly positive divergence. This has been achieved by working in the horizon-adapted coordinate system. In this work, we show that the local entropy production through the divergence of the entropy current is covariant under affine reparametrizations that leave the gauge of horizon-adapted coordinates invariant. We explicitly derive a formula for how the entropy current transforms under such coordinate transformations. This extends the analysis of \cite{Bhattacharyya:2022njk} [arXiv:2204.08447] for arbitrary diffeomorphism invariant theories of gravity non-minimally coupled to matter fields. We also study the Iyer-Wald ambiguities of the covariant phase formalism that generically plague the components of the entropy current.}

\begin{document}
\maketitle
\flushbottom


\newpage

\section{Introduction}

In modern theoretical physics, black holes serve as the best probe for understanding the mysteries of Quantum gravity. Black holes are an interesting class of solutions to Einstein's General Relativity (GR) to which one can associate thermodynamic notions like energy, temperature and, crucially, entropy given by the area of the horizon \cite{PhysRevLett.26.1344,Bardeen:1973gs,PhysRevD.7.2333,Hawking:1975vcx}. Einstein's GR is not UV complete. Effective Field Theory tells us that any UV complete theory of Quantum gravity in the low energy limit reduces to Einstein's GR with corrections coming from higher derivative interactions and couplings \cite{Donoghue:1994dn}. Since we expect to find a consistent statistical picture of thermodynamics for black holes, it is of interest to study whether the laws of black hole thermodynamics continue to hold for such higher derivative theories beyond GR \cite{PhysRevD.48.R3427,Iyer:1994ys,Jacobson:1993xs,Jacobson:1993vj,Jacobson:1995uq,Sarkar:2013swa,Bhattacharjee:2015yaa}\footnote{For a review, see \cite{Wall:2018ydq, Sarkar:2019xfd}.}.

\medskip

Iyer and Wald constructed an entropy functional that satisfied the first law of black hole thermodynamics for arbitrary diffeomorphism invariant theories of gravity \cite{PhysRevD.48.R3427,Iyer:1994ys}. Using covariant phase space methods \cite{10.1063/1.528801}, they worked out the modifications to the area law in the form of \textit{Iyer-Wald entropy} that now satisfied the first law by construction. Since the first law is a statement about equilibrium, the Iyer-Wald entropy functional was unambiguously defined only for stationary Killing horizons. It suffered from JKM ambiguities arising from the covariant phase space formalism for dynamical situations \cite{Jacobson:1993vj}. This is expected because the second law is a statement about dynamics, and these higher derivative terms modify Einstein's Equations of Motion (EoM), which is crucial for the proof of a second law.

\medskip

Wall \cite{Wall:2015raa} \footnote{Wall's algorithm has been extended to the non-linear order in \cite{Hollands:2022fkn,Davies:2023qaa,Davies:2024fut}.} fixed the JKM ambiguities in such a way that his entropy functional satisfied a second law up to linear order in dynamics. The proof of the second law was constructed in a dynamical situation where the black hole ``is settling down to equilibrium". Thus, one is perturbatively close to the equilibrium configuration (with a Killing horizon) at any point of evolution. Within this setup, the proof amounted to establishing that the entropy functional increases in time up to linear order in dynamics. 

\medskip

Following the lead of \cite{Wall:2015raa}, recently in \cite{Bhattacharyya:2021jhr,Biswas:2022grc}, the authors construct an ultra-local proof of the second law for arbitrary diffeomorphism invariant theories of gravity up to linear order in the dynamics \footnote{As argued in \cite{Adami:2021kvx}, it is reasonable to expect a local formulation of the laws of thermodynamics arising from diffeomorphism invariance.}. In its strongest form, the second law could be expressed through an \textit{entropy current} with positive divergence. This entropy current on the black hole horizon has not only a time component\footnote{\label{foot:noncov} Here, the ``time'' component is the component along the null generators of the horizon. Thus, the time is within quotes because it is a null coordinate on the horizon. Though we together call the time and spatial components a current, this is a non-covariant current. These non-covariances arise from the fact that the metric structure degenerates on the horizon. We will precisely quantify these non-covariances in this paper.} that is just Wall's modifications to Iyer-Wald entropy, but also crucially spatial components that capture the entropy re-distribution along the spatial slices of the horizon. 

\medskip

Since dynamical situations are complicated even within the purview of GR, we often resort to a choice of gauge that simplifies our analysis. \cite{Bhattacharya:2019qal,Bhattacharyya:2021jhr,Biswas:2022grc} worked with a horizon-adapted choice of coordinates detailed in \cite{Bhattacharyya:2016xfs} to establish a local entropy current on the horizon. This essentially amounts to choosing a specific slicing for the horizon. Let us now consider the following situation. Some matter falls into the horizon of an initially stationary black hole. The black hole finally settles down to a stationary solution. In this process \footnote{The Physical process version of the first law is very closely to this setup of second law \cite{Gao:2001ut,Amsel:2007mh,Chatterjee:2011wj,Bhattacharjee:2014eea,Mishra:2017sqs}.}, certain regions of the horizon might be closer to equilibrium than other regions \cite{Bhattacharya:2019qal}. This physics is captured by a flow of entropy current between the different sub-regions of the horizon. One would expect that the local entropy production should not be affected by the choice of slicing of the horizon. If the entropy production depends on the choice, then we could, in principle, choose a particular slicing with a ``sink'' where entropy decreases locally. This would not be desirable for an ultra-local version of the second law\cite{Bhattacharya:2019qal,Bhattacharyya:2021jhr,Biswas:2022grc}.  

\medskip

We aim to address this question of coordinate independence in this paper. To linearized order in dynamics of amplitude, entropy production is directly given by the divergence of the entropy current \footnote{Recently in \cite{Davies:2023qaa}, the authors have explored a non-perturbative formulation of the second law in effective field theories.}. By construction, this entropy current is intricately tied to the horizon-adapted choice of coordinates \cite{Bhattacharya:2019qal,Bhattacharyya:2021jhr}. Following up on some initial observations in \cite{Bhattacharyya:2022njk}, we will show that the entropy production in black holes of arbitrary diffeomorphism invariant theories admits a symmetry under coordinate transformations that preserve the form of the horizon-adapted gauge. This fact was explicitly verified for the particular case of Gauss-Bonnet theory by brute force computation in \cite{Bhattacharyya:2022njk}. The non-trivial part of our analysis is the fact that the components of the codimension one entropy current ``vector'' are defined on a null manifold \footnote{As mentioned in footnote \ref{foot:noncov}, though we loosely call it the codimension one current, it should be understood that this is a ``vector'' defined on a degenerate surface.}. Thus, these components do not transform covariantly under coordinate transformations. Despite these complications, we w
ill show that local entropy production remains unaffected under a change of slicing of the horizon.

\medskip

The symmetries that preserve the horizon-adapted gauge are affine reparametrizations of the null generators of the horizon \cite{Bhattacharyya:2022njk}. The transformation of the components of the entropy current ``vector'' will involve parts that are covariant, which are expected from the covariant transformation of a vector, and additional parts that are non-covariant. The non-covariant pieces can be interpreted as arising due to ambiguities in the definition of the entropy current. Due to the dynamical situation, we are interested in, the ambiguities of the covariant phase space formalism become non-trivial \cite{Jacobson:1993vj}. In this work, we study the Iyer-Wald ambiguities of the covariant phase space formalism \cite{Jacobson:1993vj} that generically plague the components of the entropy current constructed in \cite{Bhattacharyya:2021jhr}.

\medskip

This paper is organized as follows. In Sec. \ref{sec:sec2}, we give details of the statement of the problem and summarize the main results of our paper. These are: a formula for the transformation of the components of the entropy under affine reparametrizations that relate the different classes of horizon-adapted coordinates and the precise origin of ambiguities in the components of the entropy current. In Sec. \ref{sec:jvjitransfgen}, we give a formal proof of the gauge covariance of entropy production up to linear order in perturbations under affine reparametrizations of the null generators. In Sec. \ref{sec:ambcurrent}, we relate the non-covariances in the transformation of the components of the entropy current with the Iyer-Wald ambiguities of the covariant phase space formalism. We will also analyze the ambiguities in the physical process version of the first law between arbitrary time slices of the horizon. In Sec. \ref{sec:verproof}, we explicitly verify the transformation formula of the components of the entropy current for explicit diffeomorphism invariant theories of gravity. In Sec. \ref{sec:nonlinear}, we look at Gauss-Bonnet theory and study the gauge covariance to non-linear order in perturbations. We finally conclude with some discussions of our results in Sec. \ref{sec:discussions}. The details of the computation are made explicit in various appendices.

\section{A symmetry of local entropy production}
\label{sec:sec2}

In this section, we lay out the results of the main calculation by stating the problem. We will also briefly introduce the elements that will be crucial for the following sections. The basic setup and the construction of a local entropy current on the horizon are given in Appendix \ref{ap:basicsetup}. For more details, one must consult section 2 of \cite{Bhattacharyya:2021jhr}.

\subsection{Statement of the problem and summary of results}

\begin{figure}[h!]
    \centering
    \includegraphics[width = 13 cm, height = 7 cm]{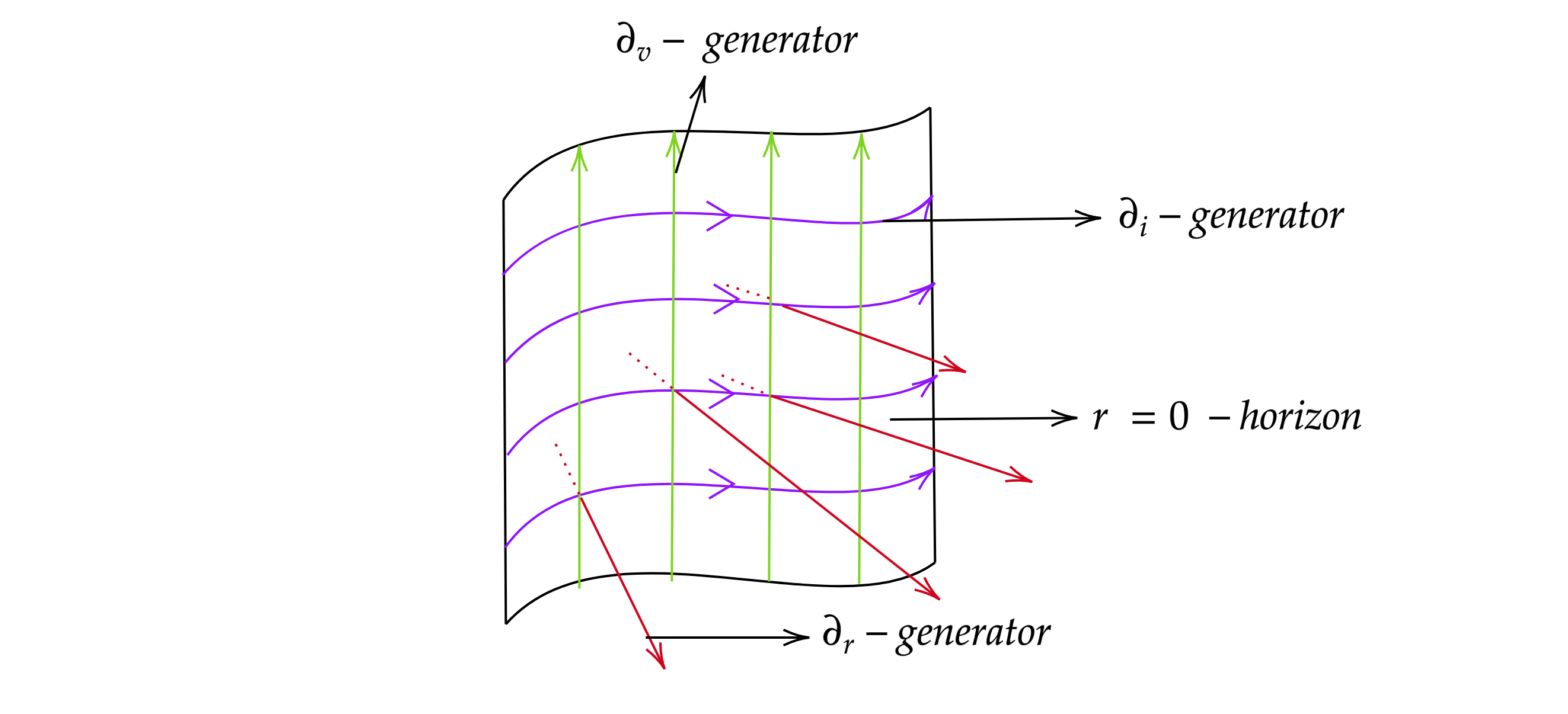}
    \caption{Schematic representation of the horizon-adapted coordinates}
    \label{fig:horizon}
\end{figure}

As mentioned in the introduction, we would be interested in the physical situation where the black hole is ``settling down to equilibrium''. Thus, in the far future, the dynamical horizon becomes a Killing horizon that satisfies the Zeroth law of black hole mechanics \cite{Bhattacharyya:2022nqa,Hollands:2022ajj,Davies:2024fut}. This dynamical situation is effectively captured by the horizon-adapted gauge of \cite{Bhattacharyya:2016xfs,Bhattacharya:2019qal,Bhattacharyya:2021jhr}. In this gauge, the spacetime of any $D$ dimensional dynamical black solution is spanned by coordinates $y^{\mu}=\{v,r,x^i\}$ with the horizon located at $r=0$. The coordinate $v$ is chosen to be an affine parameter along the null generators of the horizon $\partial_v$. Along the constant $v$ slices on the horizon, we have the spatial coordinates $x^i$ spanned by the spatial tangents $\partial_i$ of the horizon. $r$ is an affine parameter along the null geodesics generated by $\partial_r$, which make specific angles with $\partial_v$ and $\partial_i$. Thus, the metric is given by
\begin{equation}\label{eq:metric1}
    ds^2 = 2 dv \, dr - r^2 \, X(r,v,x^i) \, dv^2 + 2 r \, \omega_i(r,v,x^i) \, dv \, dx^i + h_{ij}(r,v,x^i) \, dx^i \, dx^j \, .
\end{equation}
This metric is schematically given by the figure \ref{fig:horizon}. The explicit examples of Schwarzschild and Kerr black holes expressed in this form can be found in the Appendix of \cite{Dhivakar:2023zqn}. The dynamics are encoded in the metric functions $X,\omega_i,h_{ij}$ in the form of fluctuation atop an equilibrium solution given by $g_{\mu\nu}=g^{eq}_{\mu\nu}+\epsilon \, \delta g_{\mu\nu}$. Here $g^{eq}_{\mu\nu}$ is the equilibrium solution, and $\epsilon \ll 1$ is a small parameter characterizing the deviations from an equilibrium solution. 

\medskip

Consider the most general diffeomorphism invariant theory of gravity given by the following Lagrangian
\begin{equation}\label{eq:lag1}
    \mathcal{L} = \mathcal{L}\left(g_{\mu\nu},R_{\mu\nu\rho\sigma},D_{\alpha_1}R_{\mu\nu\rho\sigma},\dots,\phi,D_{\alpha_1}\phi,\dots,F_{\mu\nu},D_{\alpha_1}F_{\mu\nu},\dots \right) \, .
\end{equation}
Here $D_{\mu}$ denotes the covariant derivative with respect to the full metric, $\phi$ denotes a scalar field, and $F_{\mu\nu}$ denotes the $U(1)$ gauge field strength tensor. Using constraints from diffeomorphism invariance and a particular symmetry of the Killing horizon of $g^{eq}_{\mu\nu}$, the authors of \cite{Bhattacharya:2019qal,Bhattacharyya:2021jhr,Biswas:2022grc} have shown that the equations of motion projected on the horizon along the null generators take a universal form at linear order, given by
\begin{equation}\label{eq:keyeqn}
    E_{vv}|_{r=0} = \partial_v \left( \dfrac{1}{\sqrt{h}} \partial_v \left( \sqrt{h} \, \mathcal{J}^v \right) + \dfrac{1}{\sqrt{h}}\partial_i \left( \sqrt{h} \, \mathcal{J}^i \right) \right) + \mathcal{O}(\epsilon^2) \, .
\end{equation}
($h=\text{det}(h_{ij})$) This special form of the $E_{vv}$ will lead to a local second law at the linear order
\begin{equation}\label{eq:localseclaw}
    \dfrac{1}{\sqrt{h}} \partial_v \left( \sqrt{h} \, \mathcal{J}^v \right) + \dfrac{1}{\sqrt{h}}\partial_i \left( \sqrt{h} \, \mathcal{J}^i \right) \geq 0 ~~ \text{upto} ~ \mathcal{O}(\epsilon^2) \, ,
\end{equation}
as explained above eq.\eqref{eq:aplocseclaw}.

\medskip

The structure of $E_{vv}$ crucially depends on the choice of slicing in eq.\eqref{eq:metric1} used. We ask the question as to what happens to the structure under a set of transformations that change the slicing of the horizon but keep the gauge intact. In other words, we are interested in understanding the covariance under gauge preserving transformations. This covariance is important to establish the robustness of the local entropy production at all regions of the horizon. Affine reparametrizations of the form
\begin{equation}\label{eq:repara}
    v \to \tau = v \, e^{-\zeta(\Vec{x})}\, , ~~~~ x^a \to y^a = x^a \, ,
\end{equation}
employ a change of coordinates $(v,r,x^a)$ to $(\tau,\rho, y^a)$ that leave the form of the metric eq.\eqref{eq:metric1} invariant \footnote{As we will detail in section \ref{ssec:difhorcoord}, one also has to do a transformation of $r$ to ensure that the form eq.\eqref{eq:metric1} is intact away from the horizon $r=0$.} \cite{Bhattacharyya:2022njk}. Thus, it is crucial to study the transformation of entropy currents under eq.\eqref{eq:repara} to see if local entropy production is indeed preserved.

\medskip

The components of the entropy current expressed through the structure of $E_{vv}$  eq.\eqref{eq:keyeqn} are also ambiguous in the following sense. Suppose we have the quantities $\mathcal{J}^v_{\text{amb}}$ and $\mathcal{J}^i_{\text{amb}}$ satisfying
\begin{equation}\label{eq:ambincurrent}
    \dfrac{1}{\sqrt{h}} \partial_v \left( \sqrt{h} \, \mathcal{J}^v_{\text{amb}} \right) + \dfrac{1}{\sqrt{h}}\partial_i \left( \sqrt{h} \, \mathcal{J}^i_{\text{amb}} \right) = \mathcal{O}(\epsilon^2) \, ,
\end{equation}
then one can always add these currents to eq.\eqref{eq:keyeqn} without changing $E_{vv}$ (i.e.)
\begin{equation}\label{eq:ambinevv}
    E_{vv}|_{r=0} = \partial_v \left( \dfrac{1}{\sqrt{h}} \partial_v \left( \sqrt{h} \, \left(\mathcal{J}^v + \mathcal{J}^v_{\text{amb}} \right) \right) + \dfrac{1}{\sqrt{h}}\partial_i \left( \sqrt{h} \, \left(\mathcal{J}^i + \mathcal{J}^i_{\text{amb}} \right) \right) \right) + \mathcal{O}(\epsilon^2) \, .
\end{equation}
Since we are relating the divergence of the current to local entropy production as in eq.\eqref{eq:localseclaw}, the components of the current contain information of entropy redistribution. Thus, it is important to study the origin of these ambiguities and quantify them appropriately.

\medskip

In this paper, we would like to address these questions. Given these details, the main results of our paper are as follows:
\begin{enumerate}
    \item The components of entropy current $\mathcal{J}^v$ and $\mathcal{J}^i$ transform under eq.\eqref{eq:repara} in the following form on the horizon $r=0$:
    \begin{equation}\label{eq:jvjitransf}
    \dfrac{1}{\sqrt{h}}\partial_v \left(\sqrt{h} \mathcal{J}^v\right) + \nabla_i \mathcal{J}^i = e^{-\zeta} \left[ \dfrac{1}{\sqrt{\widetilde{h}}} \partial_{\tau} \left( \sqrt{\widetilde{h}} \widetilde{\mathcal{J}}^{\tau} \right) + \widetilde{\nabla}_i \widetilde{\mathcal{J}}^i \right] + \mathcal{O}(\epsilon^2) \,  .
     \end{equation}
     Here $\widetilde{\mathcal{J}}^{\tau}$ and $\widetilde{\mathcal{J}}^a$ \footnote{We will use tilde to denote quantities in the $(\rho,\tau,y^i)$ coordinates.} are the components of the entropy current expressed in the $(\rho,\tau,y^i)$ coordinates and are explicitly given by
     \begin{align*}
     \fbox{
 \addtolength{\linewidth}{-2\fboxsep}%
 \addtolength{\linewidth}{-3\fboxrule}%
 \begin{minipage}{\linewidth}
  \begin{equation}\label{eq:currentres}
       \begin{split}
        \mathcal{J}^v &= \widetilde{\mathcal{J}}^{\tau} + \tau \xi_i \widetilde{\mathcal{J}}^i - 2 \tau^2 \xi_i \xi_j \partial_{\tau} \widetilde{L}^{ij}_{(1)} + 2 \tau \widetilde{\nabla}_i \left( \xi_j \widetilde{L}^{ij}_{(1)} \right) - 2 \tau \xi_i \xi_j \widetilde{L}^{ij}_{(1)} \, , \\
        \mathcal{J}^i &= e^{-\zeta} \widetilde{J}^i - 2 e^{-\zeta} (1+ \tau \partial_{\tau}) (\xi_j \widetilde{L}^{ij}_{(1)}) \, ,
    \end{split}
  \end{equation}
 \end{minipage}
}
\end{align*}
     where $\xi_i = \partial_i \zeta$ and $\widetilde{L}_{ij}$ is determined from the Lagrangian $\mathcal{L}$ of eq.\eqref{eq:lag1} as
     \begin{equation*}
         \frac{\partial\mathcal{L}}{\partial \tilde R_{\rho i \rho j}} \equiv \tilde E_R^{\rho i \rho j} \equiv \partial_\tau \widetilde{L}_{(1)}^{ij} + O(\epsilon^2) 
     \end{equation*}
     From eq.\eqref{eq:jvjitransf}, we can see that the local entropy production of eq.\eqref{eq:localseclaw} is unaffected. We show this in section \ref{sec:jvjitransfgen}.
    \item  The extraction of $\mathcal{J}^v$ and $\mathcal{J}^i$ following eq. \eqref{eq:keyeqn} is ambiguous up to terms of the form
    \begin{align*}
     \fbox{
 \addtolength{\linewidth}{-2\fboxsep}%
 \addtolength{\linewidth}{-3\fboxrule}%
 \begin{minipage}{\linewidth}
  \begin{equation}\label{eq:ambresult}
       \begin{split}
        \mathcal{J}^v &\rightarrow \mathcal{J}^v + \nabla_i C^{vi} \, , \\
        \mathcal{J}^i &\rightarrow \mathcal{J}^i - \dfrac{1}{\sqrt{h}} \partial_v \left( \sqrt{h} \, C^{vi}\right) + \nabla_j C^{ij} \, ,
    \end{split}
  \end{equation}
 \end{minipage}
}
\end{align*}
where $C^{vi}$ and $C^{ij}$ are explicitly determined in terms of the ambiguities in the symplectic potential $\Theta^{\mu}$ and Iyer-Wald Noether Charge $Q^{\mu\nu}$ derived from the Lagrangian $\mathcal{L}$ of eq.\eqref{eq:lag1}. Iyer-Wald ambiguities can occur in $\mathcal{L}, \Theta^\mu$ and $Q^{\mu\nu}$ as $\mathcal{L} \rightarrow \mathcal{L} + D_\rho M^\rho, \Theta^\mu \rightarrow \Theta\mu + D_\nu Y^{\mu\nu}$and $Q^{\mu\nu} \rightarrow Q^{\mu\nu} + D_\rho Z^{\mu\nu\rho}$. Of these, only the $Y$-type and $Z$-type ambiguities contribute to $C^{vi}$ and $C^{ij}$. This is derived in section \ref{sec:ambcurrent}.
    
\end{enumerate}

These form the main results of our paper. At this point, we mention that a special case of eq.\eqref{eq:currentres} and eq.\eqref{eq:ambresult} has been considered in \cite{Hollands:2022fkn}. The existence of $\widetilde{L}^{ij}_{(1)}$ has not been discussed before. This hints at a richer structure, and its existence is a consequence of working on a null horizon with a degenerate structure. We will comment on the comparison to \cite{Hollands:2022fkn} towards the end of section \ref{sec:jvjitransfgen} and \ref{sec:ambcurrent}.

The basic setup of the construction of an entropy current on the horizon is reviewed in Appendix \ref{ap:basicsetup}. We will now study how these affine reparametrizations of eq.\eqref{eq:repara} relate to the various choices of horizon-adapted coordinates. This will be crucial to answer the question of whether entropy production is covariant under such transformations.

\subsection{The various choices of horizon-adapted coordinates}
\label{ssec:difhorcoord}

To study the gauge covariance of local entropy production as explicated in eq.\eqref{eq:jvjitransf}, we must first study the precise relation between the different coordinate systems related by an affine reparametrization of the form eq.\eqref{eq:repara}. We follow the analysis of section 3.1 of \cite{Bhattacharyya:2022njk}. This affine reparametrization leaves the form of the metric eq.\eqref{eq:metric1} invariant. Eq.\eqref{eq:repara} is, in fact, a residual symmetry of the horizon-adapted coordinates as reviewed in eq.\eqref{eq:genvtransf}.  The transformed metric is thus written as 
\begin{equation}\label{eq:metric2}
    ds^2 = 2 d\tau \, d\rho - \rho^2 \, \widetilde{X}(\rho,\tau,y^j) \, d\tau^2 + 2 \rho \, \widetilde{\omega}(\rho,\tau,y^j) \, d\tau \, dy^i + \widetilde{h}_{ij}(\rho,\tau,y^k) \, dy^i \, dy^j \, .
\end{equation}
The transformation in eq.\eqref{eq:repara} is only valid on the horizon. To obtain the transformation, if one moves away from the horizon, the $r$ coordinate must be changed so that the angle between the tangents along the constant $\{\tau,y^a\}$ and the coordinate vector is kept constant. This leads to the redefinition of $r$ and the correction of coordinate transformation as one moves away from the horizon.
The full coordinates transformations are as follows:
\begin{equation}\label{eq:coordtransf}
    \begin{split}
        r&=\rho  e^{-\zeta } \left(1+\sum _{n=1}(\rho  \tau)^n R_n (\rho  \tau, y) \right) \, ,\\
        v&=\tau  e^{\zeta } \left(1+\sum _{n=1} (\rho  \tau)^nV_n (\rho  \tau, y)\right) \, ,\\
        x^a&=y^a+\sum _{n=1} (\rho  \tau)^nZ_n^a (\rho  \tau, y)  \, .
    \end{split}
\end{equation}

\medskip

Since we will be dealing with structures on the horizon, it suffices to have these expansion coefficients $(R_n, V_n, Z^a_n)$ up to the $n=1$ term. By imposing $D$ gauge conditions $\{g_{\rho\rho}=0\,g_{\rho\tau}=0\,g_{\rho a}=0\}$ their expressions can be derived as follows \cite{Bhattacharyya:2022njk}:
\begin{align}
    Z_1^i &=-\widetilde{h}^{ij}\xi _j \, , &
        V_1 &= -\frac{1}{2} \xi _i \xi _j \widetilde{h}^{ij} \, , &
        R_1 &= \frac{1}{2} \left(h^{ij} \omega _i \xi _j + \tau  \xi _i \xi _j \widetilde{K}^{ij}\right) \, .
\end{align}
Here, $\xi_i= \partial_i \zeta (x^a)$. Using these expressions of the expansion coefficients, the change of the metric coefficients $(X,\omega_i,h_{ij})$ up to the required orders of $\rho$ can be derived to be:
\begin{equation}\label{eq:metrictransf}
\begin{split}
        \widetilde{\omega}_i&= {\omega }_i-2 \xi _i+2\tau  \xi _k \widetilde{h}^{jk}\widetilde{K}_{ij}+\mathcal{O}(\rho ) \, , \\ 
        \widetilde{X} &=  X +\omega _i \xi _j \widetilde{h}^{ij}-\xi _i \xi _j \widetilde{h}^{ij} -\tau \left(\xi _i \widetilde{h}^{ij} \partial_\tau \omega _j-2 \omega _i \xi _j \widetilde{K}^{ij}+2 \xi _i \xi _j \widetilde{K}^{ij}\right)\\ &-\tau ^2\left(\xi _i \xi _l \widetilde{h}^{kl} \widetilde{h}^{ij}\partial_\tau \widetilde{K}_{jk}\right) + \mathcal{O}(\rho) \, , \\    
       \widetilde{ h}_{ij} &={h}_{ij}-\rho \tau \left(\xi _i \omega _j+\omega _i \xi _j-2 \xi _i \xi _j-2 \xi _{ij}+\xi _b\widetilde{h}^{ab}\widetilde{\partial }_i\widetilde{h}_{ja}+\xi _b\widetilde{h}^{ab}\widetilde{\partial }_j\widetilde{h}_{ia}\right)+O (\rho ^2) \, . \\  
\end{split}
\end{equation}
The basis vectors in the $(r,v,x^a)$ coordinates can be expressed in terms of the $(\rho,\tau,y^a)$ coordinates and the fundamental structures $(\widetilde{X},\widetilde{\omega }_i,\widetilde{h}_{i j})$ as
\begin{equation}\label{eq:basistransf}
    \begin{split}
        \partial _r&=e^{\zeta }\left(\partial _{\rho }-\frac{1}{2}\tau ^2\xi _a\xi _b\widetilde{h}^{a b}\partial _{\tau }+\tau\xi _a\widetilde{h}^{a b}\widetilde{\partial _b}\right) + O (\rho) \, , \\
        \partial_v &= e^{-\zeta} \partial_\tau + O (\rho) \, , \\
        \partial_{x^i} &= \widetilde{\partial}_{y^i} - \tau ~\xi_i ~\partial_\tau + \mathcal{O}(\rho) \, .
    \end{split}
\end{equation}
The details of how various quantities transform under a reparametrization of the form eq.\eqref{eq:repara} are given in Appendix \ref{ap:reparadetails}. These structures will be important when we study how the components of the entropy current transform in the subsequent section \ref{sec:jvjitransfgen}.

\section{Proof of gauge covariance of the entropy current}
\label{sec:jvjitransfgen}

In this section, we will set out to prove the gauge covariance of local entropy production for arbitrary diffeomorphism invariant theories of gravity as predicted by eq.\eqref{eq:jvjitransf}. This was checked for a particular case of Gauss-Bonnet theory in \cite{Bhattacharyya:2022njk}. We will generalize their result for arbitrary diffeomorphism invariant theories of gravity non-minimally coupled to matter fields of the form in eq.\eqref{eq:lag1}. The currents $\mathcal{J}^v$ and $\mathcal{J}^i$ are quantities defined on a null surface; consequently, they are non-covariant structures. Thus, they must transform highly non-trivially for eq.\eqref{eq:jvjitransf} to be satisfied. We will explicitly derive the first result of our paper given in eq.\eqref{eq:currentres}. The details of the calculations are presented in Appendix \ref{ap:detailscomp}.

\medskip

\begin{figure}
    \centering
    \includegraphics[width = 15 cm, height = 9 cm]{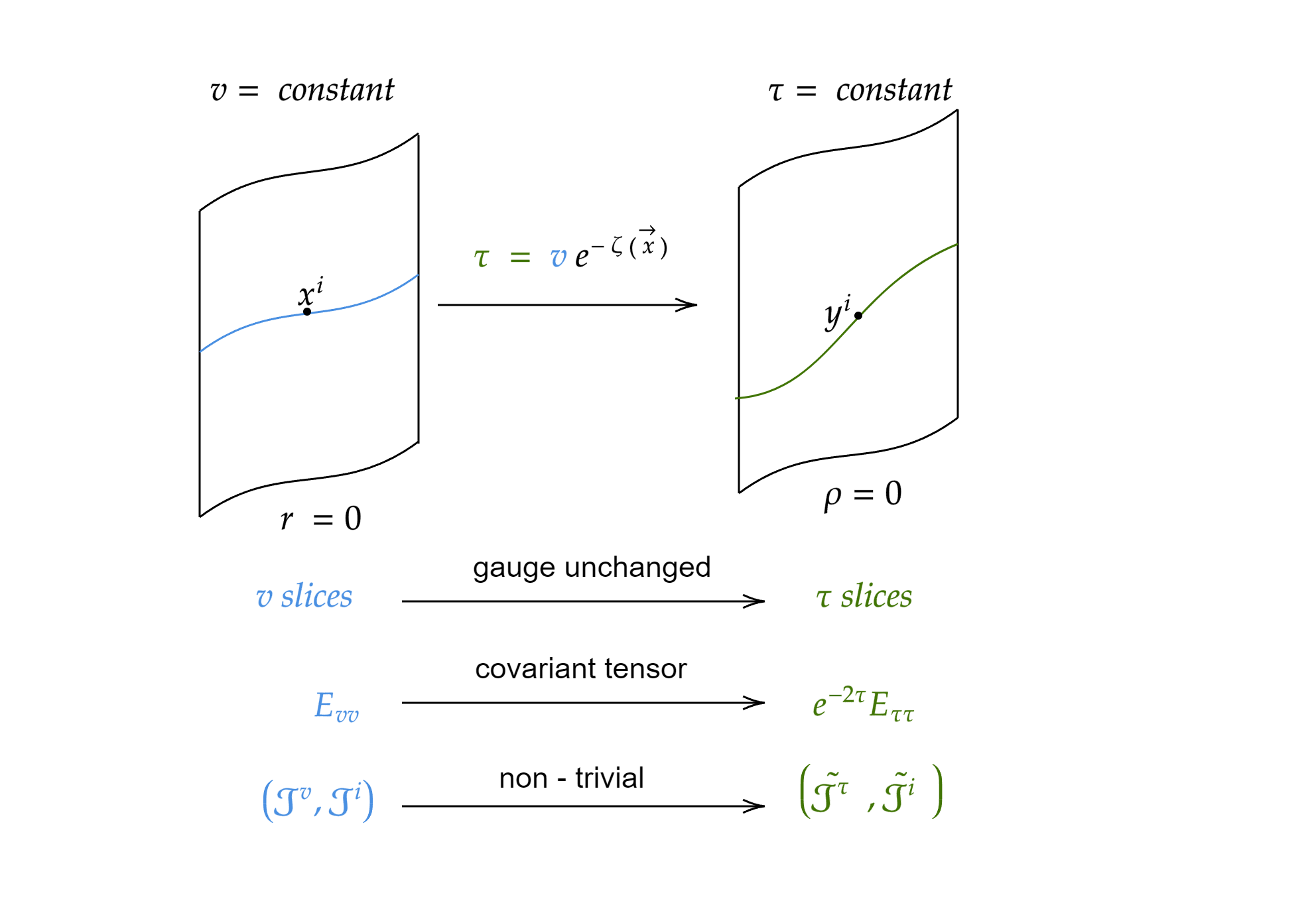}
    \caption{Different slicings of the horizon-adapted coordinates}
    \label{fig:slicing}
\end{figure}

We will start with the key equation eq.\eqref{eq:mainrel} that goes into the proof of the existence of the entropy current for arbitrary diffeomorphism invariant theories of gravity \cite{Bhattacharyya:2021jhr}:
\begin{equation}\label{eq:mainevvrel}
    2v \, E_{vv}|_{r=0} = \left(-\Theta^r + D_{\mu}Q^{r\mu} \right)|_{r=0} \, .
\end{equation}
$\Theta^r$ and $Q^{r\mu}$ evaluated on the horizon $r=0$ take the form of eq.\eqref{eq:thetaqfinal}:
\begin{equation}\label{eq:thetaqon}
    \begin{split}
        \Theta^r|_{r=0} &= (1+v\partial_v) \mathcal{A}_{(1)} + v\partial^2_v \mathcal{B}_{(0)} + \mathcal{O}(\epsilon^2) \, , \\
        Q^{r\mu} &= \mathcal{Q}^{r\mu}+v \, W^{r\mu}_v \, .
    \end{split}
\end{equation}
This results in eq.\eqref{eq:evvapfinal} and thus, we have proved eq.\eqref{eq:keyeqn} with the components of the current given by
\begin{equation}\label{eq:jvjidef}
    \mathcal{J}^v = -\dfrac{1}{2}\left( \mathcal{Q}^{rv} + \mathcal{B}_{(0)} \right) \, , ~~~~~~ \mathcal{J}^i = -\dfrac{1}{2} \left( \mathcal{Q}^{ri} -  J^i_{(1)} \right) \, .
\end{equation}
The subscripts in the above expressions denote the boost weight (eq.\eqref{eq:boostwt}) of various quantities. Under an affine reparametrization of the form eq.\eqref{eq:repara},
\begin{equation}\label{eq:repara2}
     v \to \tau = v \, e^{-\zeta(\Vec{x})}\, , ~~~~ x^a \to y^a = x^a \, ,
\end{equation}
\begin{equation}\label{eq:evvtrans}
    E_{vv} = e^{-2\zeta} E_{\tau \tau} \, ,
\end{equation}
as EoM of any diffeomorphism invariant Lagrangian is covariant. The result eq.\eqref{eq:jvjitransf} given by
\begin{equation}\label{eq:jvjitransf2}
    \dfrac{1}{\sqrt{h}}\partial_v \left(\sqrt{h} \mathcal{J}^v\right) + \nabla_i \mathcal{J}^i = e^{-\zeta} \left[ \dfrac{1}{\sqrt{\widetilde{h}}} \partial_{\tau} \left( \sqrt{\widetilde{h}} \widetilde{\mathcal{J}}^{\tau} \right) + \widetilde{\nabla}_i \widetilde{\mathcal{J}}^i \right] + \mathcal{O}(\epsilon^2) \,  ,
\end{equation}
which we are trying to prove is certainly consistent with  eq.\eqref{eq:evvtrans}. But to argue that eq.\eqref{eq:evvtrans} implies eq.\eqref{eq:jvjitransf2} is non-trivial. This is because $\mathcal{J}^v$ and $\mathcal{J}^i$ are non-covariant structures according to eq.\eqref{eq:jvjidef}. The main element of our proof is that the non-covariant terms are extracted from covariant quantities like $\Theta^r$ and $Q^{r\mu}$ through eq.\eqref{eq:thetaqon}. We know how covariant quantities transform under eq.\eqref{eq:repara}, and this, in turn, will fix how the non-covariant quantities transform under eq.\eqref{eq:repara}. The main steps of the proof are summarized schematically in the figure \ref{fig:slicing}.

\subsubsection*{Structures in $Q^{r\mu}$:}

We will first work out the transformation properties of structures extracted from $Q^{r\mu}$. The Noether charge $Q^{r\mu}$ is covariant. The construction of the Noether charge in Proposition 4.1 of \cite{Iyer:1994ys} has shown that
\begin{equation}
    Q^{\mu\nu}= W^{\mu\nu\rho}\xi_\rho-2 E^{\mu\nu\alpha\beta}_R D_{[\alpha}\xi_{\beta]}+Y^{\mu\nu}+D_\alpha Z^{\mu\nu\alpha},
\end{equation}
where $Y^{\mu\nu}$ and $Z^{\mu\nu\alpha}$ are the possible ambiguities in the definition of Noether Charge. The subtleties of the ambiguities will be carefully analysed in section \ref{sec:ambcurrent}. Neglecting the ambiguities for the purposes of this section, we have
\begin{equation}\label{eq:qstruc}
    Q^{\mu\nu} = W^{\mu\nu\rho}\xi_{\rho} - 2 E^{\mu\nu\rho\sigma}_R D_{[\rho} \xi_{\sigma]} \, ,
\end{equation}
where $E^{\mu\nu\rho\sigma}_R$ is defined from the Lagrangian in eq.\eqref{eq:lag1} as \footnote{``$(\mu \nu)$'' denote the symmetrization in $\mu$ and $\nu$ indices and ``$[\mu \nu]$'' denote anti symmetrization. The convention used for symmetrization/antisymmetrization of indices is as follows: $A^{(\alpha} B^{\beta)}=({1}/{2})(A^\alpha B^\beta+ A^\beta B^\alpha$).}
\begin{equation}\label{eq:emunurhosigma}
    E^{\mu\nu\rho\sigma}_R = \dfrac{\partial \, \mathcal{L}}{\partial R_{\mu\nu\rho\sigma}} - D_{\alpha_1} \left( \dfrac{\partial \, \mathcal{L}}{\partial D_{\alpha_1} R_{\mu\nu\rho\sigma}} \right) + D_{(\alpha_1} D_{\alpha_2)} \left( \dfrac{\partial \, \mathcal{L}}{\partial D_{(\alpha_1}D_{\alpha_2)} R_{\mu\nu\rho\sigma}}\right) + \dots  \, ,
\end{equation}
and $\xi^{\mu}$ is the Killing vector eq.\eqref{eq:killing}. Since eq.\eqref{eq:qstruc} is linear in $\xi^{\mu}$, the structure of $Q^{r\mu}$ on the horizon $r=0$ is given by eq.\eqref{eq:thetaqfinal}
\begin{equation}\label{eq:qmunuexpr}
    Q^{r\mu}|_{r=0} = \mathcal{Q}^{r\mu}+v \, W^{r\mu}_v \, ,
\end{equation}
where the explicit expressions are
\begin{equation}\label{eq:qwexpr}
    \begin{split}
        \mathcal{Q}^{rv} &= 4 E^{rvrv}_R \, , ~~~~~~ \mathcal{Q}^{ri} = 4 E^{rirv}_R \, , \\
        W^{rv}_v &= W^{rvr} - 2 E^{rirv}_R \omega_i \, , \\
        W^{ri}_v &= W^{rir} - 2 E^{rirj}_R \omega_j = \partial_v J^i_{(1)} + \mathcal{O}(\epsilon^2) \, .
    \end{split}
\end{equation}
In the expression of $W^{ri}_v$, we have chosen to express a boost weight $+2$ quantity as $\partial_v J^i_{(1)}$ up to terms that are $\mathcal{O}(\epsilon^2)$ according to eq.\eqref{eq:result1}. This structure of $Q^{r\mu}$ in \eqref{eq:qwexpr} makes it easier to study the transformation property of $\mathcal{Q}^{r\mu}$ and $W^{r\mu}_v$. We will see that these transformation properties will fix the transformation of $\mathcal{A}_{(1)}$ and subsequently $\mathcal{B}_{(0)}$.

\medskip

Since $\mathcal{L}$ of eq.\eqref{eq:lag1} is a diffeomorphism invariant Lagrangian, $E^{\mu\nu\rho\sigma}_R$ given by eq.\eqref{eq:emunurhosigma} is a covariant tensor. Thus, under a reparametrization of eq.\eqref{eq:repara}, using the metric transformation eq.\eqref{eq:metrictransf}, we can straightforwardly work out how the various quantities in eq.\eqref{eq:qwexpr} transform. For instance, the transformation of $\mathcal{Q}^{ri}$ and $W^{ri}_v$ is given by
\begin{equation}\label{eq:qriw}
    \begin{split}
        \mathcal{Q}^{ri}  &= e^{-\zeta} \widetilde{\mathcal{Q}}^{\rho i } + 4 e^{-\zeta} \tau \xi_b \widetilde{E}^{\rho i \rho b}_R \, , \\
        W^{ri}_v &= e^{-2\zeta} \widetilde{W}^{\rho i}_{\tau} - 4 e^{-2\zeta} \xi_b \widetilde{E}^{\rho i \rho b}_R \, .
    \end{split}
\end{equation}
This is checked in eq.\eqref{eq:apqritransf} and eq.\eqref{eq:apwrivtransf}. As a sanity check, one can see that the eq.\eqref{eq:qriw} ensures that $Q^{ri}$ transforms covariantly as expected (i.e.) it transforms as
\begin{equation}
    Q^{ri} = \dfrac{\partial r}{\partial \rho}\dfrac{\partial x^i}{\partial y^a} \widetilde{Q}^{\rho a} = e^{-\zeta} \widetilde{Q}^{\rho i} \, .
\end{equation}
From the transformation of $W^{ri}_v$ of eq.\eqref{eq:qriw}, we can derive the transformation of $J^i_{(1)}$. For this, we first define a boost weight $+1$ quantity $\widetilde{L}^{ij}_{(1)}$ through the following equation:
\begin{equation}\label{eq:lij1}
    \widetilde{E}^{\rho i \rho j}_R = \partial_{\tau} \widetilde{L}^{ij}_{(1)} + \mathcal{O}(\epsilon^2) \, .
\end{equation}
This is because $E^{\rho i \rho j}_R$ is a boost weight $+2$ tensor, and we have used eq.\eqref{eq:result1}\footnote{It has been recently proved \cite{Hollands:2022fkn} that the notion of boost weight remains invariant under an affine reparametrization of the form eq.\eqref{eq:repara}.}. Thus from eq.\eqref{eq:qriw}, we get
\begin{equation}\label{eq:ji1transf}
    J^i_{(1)} = e^{-\zeta} \widetilde{J}^i_{(1)} - 4 e^{-\zeta} \xi_j \widetilde{L}^{ij}_{(1)} \, .
\end{equation}
This is checked in eq.\eqref{eq:apji1transf}. Finally, from the transformation of $\mathcal{Q}^{ri}$ of eq.\eqref{eq:qriw} and $J^i_{(1)}$ of eq.\eqref{eq:ji1transf}, we get the transformation for $\mathcal{J}^i$ of eq.\eqref{eq:jvjidef} as
\begin{equation}\label{eq:jifinaltransf}
    \mathcal{J}^i = e^{-\zeta} \widetilde{\mathcal{J}}^i - 2 e^{-\zeta} (1 + \tau \partial_{\tau})(\xi_j \widetilde{L}^{ij}_{(1)}) \, .
\end{equation}
This is checked in eq.\eqref{eq:apjitransf}.

To look at how $\mathcal{J}^v$ of eq.\eqref{eq:jvjidef} transforms, we need the transformation of $\mathcal{Q}^{rv}$ and $\mathcal{B}_{(0)}$. $\mathcal{Q}^{rv}$ of eq.\eqref{eq:qwexpr} transforms as a tensor because of eq.\eqref{eq:emunurhosigma}:
\begin{equation}\label{eq:qtilderv}
    \mathcal{Q}^{rv} = \widetilde{\mathcal{Q}}^{\rho\tau} + 2 \tau \xi_i \widetilde{\mathcal{Q}}^{\rho i} + 4 \tau^2 \xi_i \xi_j \partial_{\tau} \widetilde{L}^{ij}_{(1)} \, .
\end{equation}
This is checked in eq.\eqref{eq:apvarqrvtransf}. Similarly, one can find the transformation of $W^{rv}_v$ of eq.\eqref{eq:qwexpr} as follows
\begin{equation}\label{eq:wrvv}
    W^{rv}_v = e^{-\zeta} \widetilde{W}^{\rho\tau}_{\tau} + e^{-\zeta} \tau \xi_i \widetilde{W}^{\rho i}_{\tau} - 4 e^{-\zeta} \xi_i \widetilde{E}^{\rho i \rho \tau}_R - 4 \tau e^{-\zeta} \xi_i \xi_j \partial_{\tau} \widetilde{L}^{ij}_{(1)} \, .
\end{equation}
This is checked in eq.\eqref{eq:apwrvvtransf}. Again, as a sanity check, using eq.\eqref{eq:qtilderv} and eq.\eqref{eq:wrvv}, one can straightforwardly see that $Q^{rv}$ transforms as expected (as a tensor):
\begin{equation}\label{eq:qrvtransf}
    Q^{rv} = \widetilde{Q}^{\rho \tau} + \tau \xi_i \widetilde{Q}^{\rho i} \, .
\end{equation}
This is checked in eq.\eqref{eq:apqrvtransf}. This completes our analysis of the structures in $Q^{r\mu}$.

\subsubsection*{Structures in $\Theta^r$:}

We now focus on the structures in $\Theta^r$. This is the non-trivial part of our exercise. The transformation of $\mathcal{Q}^{rv}$ of eq.\eqref{eq:qtilderv}, $\mathcal{Q}^{ri}$ of eq.\eqref{eq:qriw}, and the transformation of $W^{rv}_v$ of eq.\eqref{eq:wrvv} completely fix the transformation of $\mathcal{A}_{(1)}$ in the structure of $\Theta^r$ given by eq.\eqref{eq:thetaqon}. This is crucially because of the identity given by eq.\eqref{eq:preevv2} (eq.(3.69) of \cite{Bhattacharyya:2021jhr}):
\begin{equation}\label{eq:a1identity}
    \mathcal{A}_{(1)} = \dfrac{1}{\sqrt{h}}\partial_v \left( \sqrt{h} \, \mathcal{Q}^{rv} \right) + \dfrac{1}{\sqrt{h}} \partial_i \left( \sqrt{h} \, \mathcal{Q}^{ri} \right) + W^{rv}_v \, .
\end{equation}
From the transformation of the quantities in the RHS of eq.\eqref{eq:a1identity}, one can determine the transformation of $\mathcal{A}_{(1)}$ as
\begin{equation}\label{eq:a1finaltransf}
    \mathcal{A}_{(1)} = e^{-\zeta} \widetilde{\mathcal{A}}_{(1)} + e^{-\zeta} \tau \xi_i \partial_{\tau} \widetilde{\mathcal{Q}}^{\rho i} + e^{-\zeta} \tau \widetilde{\nabla_i} \left( 4 \xi_j \partial_{\tau} \widetilde{L}^{ij}_{(1)} \right) + e^{-\zeta} \tau \xi_i \partial_{\tau} \widetilde{J}^i_{(1)} - 4 e^{-\zeta} \tau \xi_i \xi_j \partial_{\tau} \widetilde{L}^{ij}_{(1)} \, .
\end{equation}
This is derived in eq.\eqref{eq:apa1transf}. We now use the most important point that is central to the proof. $\Theta^r$ is covariant, and thus, it transforms covariantly under the reparametrization of eq.\eqref{eq:repara}. But $\mathcal{A}_{(1)}$ and $\mathcal{B}_{(0)}$ are not covariant. Thus, $\mathcal{A}_{(1)}$ and $\mathcal{B}_{(0)}$ must transform non-trivially to $\widetilde{\mathcal{A}}_{(1)}$ and $\widetilde{\mathcal{B}}_{(0)}$ such that the following equation holds:
 \begin{equation}\label{eq:thetartransf}
     \Theta^r |_{r=0} = \dfrac{\partial r}{\partial \rho} \widetilde{\Theta}^{\rho} |_{\rho = 0} = e^{-\zeta} \widetilde{\Theta}^{\rho} |_{\rho=0} \, .
 \end{equation}
The transformation of $\mathcal{A}_{(1)}$ fixes the transformation of $\mathcal{B}_{(0)}$ because we have to ensure that $\Theta^r$ transforms like a tensor according to eq.\eqref{eq:thetartransf}. Thus, we have:
\begin{equation}\label{eq:botransf}
    \mathcal{B}_{(0)} = \widetilde{\mathcal{B}}_{(0)} - \tau \xi_i \widetilde{\mathcal{Q}}^{\rho i} - \tau \xi_i \widetilde{J}^i_{(1)} - \tau \widetilde{\nabla}_i \left( 4 \xi_j \widetilde{L}^{ij}_{(1)}\right) + 4 \tau \xi_i \xi_j \widetilde{L}^{ij}_{(1)} \, .
\end{equation}
This is checked in eq.\eqref{eq:apbotransf}. From the transformation of $\mathcal{Q}^{rv}$ of eq.\eqref{eq:qtilderv}, and that of $\mathcal{B}_{(0)}$ given in eq.\eqref{eq:botransf}, one can straightforwardly work out how $\mathcal{J}^v$ of eq.\eqref{eq:jvjidef} transforms. $\mathcal{J}^v$ transforms as
\begin{equation}\label{eq:jvtransf}
    \mathcal{J}^v = \widetilde{\mathcal{J}}^{\tau} + \tau \xi_i \widetilde{\mathcal{J}}^i - 2 \tau^2 \xi_i \xi_j \partial_{\tau} \widetilde{L}^{ij}_{(1)} + 2 \tau \widetilde{\nabla}_i \left( \xi_j \widetilde{L}^{ij}_{(1)} \right) - 2 \tau \xi_i \xi_j \widetilde{L}^{ij}_{(1)} \, .
\end{equation}
This is derived in eq.\eqref{eq:apjvtransf}.

\medskip

Thus putting together eq.\eqref{eq:ji1transf} and eq.\eqref{eq:jvtransf}, we have the final transformation of $\mathcal{J}^v$ and $\mathcal{J}^i$ as follows:
\begin{equation}\label{eq:jvjifinaltransf}
    \begin{split}
        \mathcal{J}^v &= \widetilde{\mathcal{J}}^{\tau} + \tau \xi_i \widetilde{\mathcal{J}}^i - 2 \tau^2 \xi_i \xi_j \partial_{\tau} \widetilde{L}^{ij}_{(1)} + 2 \tau \widetilde{\nabla}_i \left( \xi_j \widetilde{L}^{ij}_{(1)} \right) - 2 \tau \xi_i \xi_j \widetilde{L}^{ij}_{(1)} \, , \\
        \mathcal{J}^i &= e^{-\zeta} \widetilde{J}^i - 2 e^{-\zeta} (1+ \tau \partial_{\tau}) (\xi_j \widetilde{L}^{ij}_{(1)}) \, .
    \end{split}
\end{equation}
Here $\widetilde{L}^{ij}_{(1)}$ is defined by eq.\eqref{eq:lij1}. This proves the first result of our paper given in eq.\eqref{eq:currentres}. We will verify the formula of eq.\eqref{eq:jvjifinaltransf} for various Lagrangians in section \ref{sec:verproof}. To understand the significance of $\widetilde{L}^{ij}_{(1)}$, notice that if you set $\widetilde{L}^{ij}_{(1)} = 0$, then $\mathcal{J}^v$ and $\mathcal{J}^i$ would transform covariantly. $\mathcal{J}^v$ would transform like a covariant vector with an upper index, and $\mathcal{J}^i$ would transform like a covariant boost weight $+1$ quantity. Thus, $\widetilde{L}^{ij}_{(1)}$ signifies a non-covariance associated with the transformation of the components of entropy current defined on the dynamical horizon. The non-trivial consequence of this analysis is that despite these non-covariances, the local entropy production remains covariant (eq.\eqref{eq:jvjitransf2}) under affine reparametrizations of the form eq.\eqref{eq:repara}.

\medskip

At this point, it is useful to comment on the relation between eq.\eqref{eq:jvjifinaltransf} and the result of eq(131) of \cite{Hollands:2022fkn} which, when translated to our notation, becomes:
\begin{equation}\label{eq:hkrres}
    \begin{split}
        \widetilde{\mathcal{J}}^{\tau} &= \mathcal{J}^v - \nabla_i B^{vi} \, , \\
        e^{-\zeta}\widetilde{\mathcal{J}}^i &= \mathcal{J}^i + \dfrac{1}{\sqrt{h}} \partial_v \left( \sqrt{h} \, B^{vi} \right) + \nabla_j B^{ij} \, .
    \end{split}
\end{equation}
Our result eq.\eqref{eq:jvjifinaltransf} expressed in the $(r,v,x^i)$ coordinates is given by
\begin{equation}\label{eq:jvjitransfrvxi2}
    \begin{split}
        \widetilde{\mathcal{J}}^{\tau} &= \mathcal{J}^v - v \xi_i \left[ \mathcal{J}^i + \dfrac{1}{\sqrt{h}} \partial_v \left( \sqrt{h} \, 2v \, \xi_j L^{ij}_{(1)} \right) \right] + \nabla_i \left( -2v \, \xi_j L^{ij}_{(1)} \right) \, , \\
        e^{-\zeta} \widetilde{\mathcal{J}}^i &= \mathcal{J}^i + \dfrac{1}{\sqrt{h}} \partial_v \left( \sqrt{h} \, 2 v \, \xi_j L^{ij}_{(1)} \right) \, .
    \end{split}
\end{equation}
However, to compare eq.\eqref{eq:jvjitransfrvxi2} with eq.\eqref{eq:hkrres}, one must set $v=\tau=0$ because eq.\eqref{eq:hkrres} has been derived for $v=\tau=0$ slice of the black hole horizon \cite{Hollands:2022fkn}. Setting $v=0$ in eq.\eqref{eq:jvjitransfrvxi2}, we get
\begin{equation}\label{eq:jvjiv0}
    \widetilde{\mathcal{J}}^{\tau} = \mathcal{J}^v \, , ~~~~~~ e^{-\zeta} \widetilde{\mathcal{J}}^i = \mathcal{J}^i +2  \xi_j L^{ij}_{(1)} \, .
\end{equation}
Comparing eq.\eqref{eq:jvjiv0} to eq.\eqref{eq:hkrres}, we see that to have a non-trivial $B^{vi}$, one must not restrict to the $v=\tau=0$ slice. Thus, in order to see the richer structure in eq.\eqref{eq:hkrres}, one must not restrict to the $v=\tau=0$ slice as we have done in eq.\eqref{eq:jvjifinaltransf}.

\medskip

We emphasize the difference between the two approaches: \cite{Hollands:2022fkn} is concerned with the gauge invariance of the definition of dynamical entropy $\mathcal{J}^v$, which in the linear order is the statement of eq.\eqref{eq:jvjiv0}. Our approach looks at the gauge covariance of entropy production through eq.\eqref{eq:jvjitransf2}. This is slightly different because in our approach, both $\mathcal{J}^v$ and $\mathcal{J}^i$ transform non-trivially to ensure that the covariant relation of eq.\eqref{eq:evvtrans} is respected. This non-triviality is due to the presence of $\widetilde{L}^{ij}_{(1)}$ in eq.\eqref{eq:jvjifinaltransf}. In the subsequent section, we will study the Iyer-Wald ambiguities in the context of entropy current, and we will show that they have a structure similar to the non-covariant terms in eq.\eqref{eq:jvjitransfrvxi2}.

\section{Iyer-Wald Ambiguities of the entropy current}
\label{sec:ambcurrent}

In this section, our aim will be to establish the second result of our paper given in eq.\eqref{eq:ambresult}. Consider the structure of $E_{vv}$ at the linear order given by eq.\eqref{eq:keyeqn}
\begin{equation}
    E_{vv} = \partial_v \left[ \dfrac{1}{\sqrt{h}}\partial_v \left(\sqrt{h} \mathcal{J}^v\right) + \nabla_i \mathcal{J}^i \right] + \mathcal{O}(\epsilon^2) \, .
\end{equation}
One can straightforwardly see that
\begin{equation}\label{eq:ambiguitymain}
    \begin{split}
        \mathcal{J}^v &\rightarrow \mathcal{J}^v + \nabla_i C^{vi} \, , \\
        \mathcal{J}^i &\rightarrow \mathcal{J}^i - \dfrac{1}{\sqrt{h}} \partial_v \left( \sqrt{h} \, C^{vi}\right) + \nabla_j C^{ij} \, ,
    \end{split}
\end{equation}
with $C^{ij} = - C^{ji}$ will ensure that $E_{vv}$ remains the same though the components of entropy current change. Thus the additional terms in eq.\eqref{eq:ambiguitymain} given by
\begin{equation}\label{eq:addambs}
    \mathcal{J}^v_{\text{amb}} = \nabla_i C^{vi} \, , ~~~~~ \mathcal{J}^i_{\text{amb}} = - \dfrac{1}{\sqrt{h}} \partial_v \left( \sqrt{h} \, C^{vi}\right) + \nabla_j C^{ij} \, ,
\end{equation}
constitute an ambiguity in the definition of the entropy currents $\mathcal{J}^v$ and $\mathcal{J}^i$. This is because eq.\eqref{eq:addambs} satisfies eq.\eqref{eq:ambincurrent}. Given the generic existence of ambiguities via eq.\eqref{eq:ambiguitymain}, it is important to understand the origin of these ambiguities from the fundamental elements. We will show that the ambiguities of the form eq.\eqref{eq:ambiguitymain} arise from the ambiguities in $\Theta^r$ and $Q^{r\mu}$ of the defining equation eq.\eqref{eq:mainevvrel}. 

\subsection{The different types of ambiguities}

We want to arrive at eq.\eqref{eq:ambiguitymain} from the first principle arguments. We will see that it arises from the Iyer-Wald ambiguities \cite{Iyer:1994ys} that are part of the definition of the Lagrangian $\mathcal{L}$, $\Theta^{\mu}$ and $Q^{\mu\nu}$. From the variation of the Lagrangian of the form eq.\eqref{eq:lag1}, we have
\begin{equation}\label{eq:varlamb}
    \delta (\sqrt{-g} \mathcal{L}) = \sqrt{-g} E^{\mu\nu} \delta g_{\mu\nu} + \sqrt{-g} G^{\mu} \delta A_{\mu} + \sqrt{-g} D_{\mu} \Theta^{\mu}[\delta g_{\alpha\beta},\delta A_{\gamma}] \, .
\end{equation}
Here $E^{\mu\nu}$ are the gravitational EoM and $G^{\mu}$ is the gauge field EoM. We see that there is a $Y-$type ambiguity in the definition of $\Theta^{\mu}$ itself, which is given by
\begin{equation}\label{eq:ytypeamb}
    \Theta^{\mu} \rightarrow \Theta^{\mu} + D_{\nu} Y^{\mu\nu} \, ,
\end{equation}
where $Y^{\mu\nu} = - Y^{\nu\mu}$. Thus, the definition of $\Theta^{\mu}$ is ambiguous up to this $Y^{\mu\nu}$. The definition of the Lagrangian $\mathcal{L}$ itself is ambiguous up to a total derivative term:
\begin{equation}\label{eq:lamb}
    \mathcal{L} \rightarrow \mathcal{L} + D_{\mu} M^{\mu}  \, .
\end{equation}
eq.\eqref{eq:lamb} doesn't affect the equations of motion but leaves an ambiguity in $\Theta^{\mu}$ given by
\begin{equation}\label{eq:mtypeamb}
    \Theta^{\mu} \rightarrow \Theta^{\mu} + \delta M^{\mu} + \dfrac{1}{2} M^{\mu} g^{\alpha \beta} \delta g_{\alpha\beta} \, .
\end{equation}
Combining eq.\eqref{eq:ytypeamb} and eq.\eqref{eq:mtypeamb}, we have
\begin{equation}\label{eq:combthetaamb}
    \Theta^{\mu}[\delta g_{\alpha\beta},\delta A_{\gamma}] \rightarrow \Theta^{\mu}[\delta g_{\alpha\beta},\delta A_{\gamma}] + D_{\nu} Y^{\mu\nu}[\delta g_{\alpha\beta},\delta A_{\gamma}] + \delta M^{\mu} + \dfrac{1}{2} M^{\mu} g^{\alpha \beta} \delta g_{\alpha\beta} \, .
\end{equation}
One can obtain the Iyer-Wald Noether charge $Q^{\mu\nu}$ from eq.\eqref{eq:varlamb} if one considers the variation of the Lagrangian under a diffeomorphism given by eq.\eqref{eq:Ldiff}. The definition of $Q^{\mu\nu}$ is through eq.\eqref{eq:jmu} \footnote{$\Lambda$ is the $U(1)$ gauge transformation. Thus one has $\delta A_{\mu} = \mathcal{L}_{\xi}A_{\mu} + D_{\mu}\Lambda$. Substituting this in $\Theta^{\mu}$ brings out a factor of $A_{\nu}\xi^{\nu} + \Lambda$. This is worked out in Section 6.1 of \cite{Biswas:2022grc} (see, for instance eq.(6.6) of \cite{Biswas:2022grc}).}:
\begin{equation}\label{eq:defqmunu}
    \Theta^{\mu}[\mathcal{L}_{\xi}g_{\mu\nu},\dots] - \xi^{\mu}\mathcal{L} = -2 E^{\mu\nu}\xi_{\nu} - G^{\mu}(A_{\nu}\xi^{\nu} + \Lambda) + D_{\nu} Q^{\mu\nu} \, .
\end{equation}
This definition of $Q^{\mu\nu}$ will automatically induce an ambiguity of the form 
\begin{equation}\label{eq:ztypeamb}
    Q^{\mu\nu} \rightarrow Q^{\mu\nu} + D_{\rho} Z^{\mu\nu\rho} ~~~~ \text{where} ~ Z^{\mu\nu\rho} = Z^{[\mu\nu\rho]} \, .
\end{equation}
$Q^{\mu\nu}$ will also inherit the ambiguities of $\Theta^{\mu}$ of eq.\eqref{eq:combthetaamb}. To find out the contributions due to the $Y-$type ambiguity of eq.\eqref{eq:ytypeamb} and the $M-$type ambiguity of eq.\eqref{eq:mtypeamb}, we start with the $\Theta^{\mu}$ in the definition of $Q^{\mu\nu}$ in eq.\eqref{eq:defqmunu}:
\begin{equation}
    \begin{split}
        \Theta^{\mu}[\mathcal{L}_{\xi}g_{\alpha\beta},\mathcal{L}_{\xi}A_{\gamma} + D_{\gamma}\Lambda] &\rightarrow \Theta^{\mu}[\mathcal{L}_{\xi}g_{\alpha\beta},\mathcal{L}_{\xi}A_{\gamma} + D_{\gamma}\Lambda] + D_{\nu} Y^{\mu\nu}[\mathcal{L}_{\xi}g_{\alpha\beta},\mathcal{L}_{\xi}A_{\gamma} + D_{\gamma}\Lambda] \\
        & ~~~~ + \mathcal{L}_{\xi} M^{\mu} + \dfrac{1}{2}M^{\mu} g^{\alpha\beta}\mathcal{L}_{\xi} g_{\alpha\beta} \\
        &\rightarrow \Theta^{\mu}[\mathcal{L}_{\xi}g_{\alpha\beta},\mathcal{L}_{\xi}A_{\gamma} + D_{\gamma}\Lambda] + D_{\nu} Y^{\mu\nu}[\mathcal{L}_{\xi}g_{\alpha\beta},\mathcal{L}_{\xi}A_{\gamma} + D_{\gamma}\Lambda] \\
        & ~~~~ + \xi^{\mu} D_{\rho} M^{\rho} + D_{\nu} \left[ M^{\mu}\xi^{\nu} - M^{\nu} \xi^{\mu} \right] \, .
    \end{split}
\end{equation}
Note that crucially, we use $\delta M^{\mu} = \mathcal{L}_{\xi}M^{\mu}$ since we consider total derivative terms that are diffeomorphism covariant and $U(1)$ gauge invariant. Thus, we will not have anything to say about Chern-Simons theories, which are diffeomorphism invariant ($U(1)$ invariant) up to total derivatives only. These theories have been recently analysed in \cite{Deo:2023vvb}. From eq.\eqref{eq:defqmunu}, we see that the contribution of eq.\eqref{eq:combthetaamb} to $Q^{\mu\nu}$ is given by 
\begin{equation}\label{eq:Qmamb}
    Q^{\mu\nu} \rightarrow Q^{\mu\nu} + Y^{\mu\nu} + M^{\mu}\xi^{\nu} - M^{\nu} \xi^{\mu} \, .
\end{equation}
Hence the combined ambiguity in $Q^{\mu\nu}$ from eq.\eqref{eq:ztypeamb} and eq.\eqref{eq:Qmamb} is given by
\begin{equation}\label{eq:combambq}
    Q^{\mu\nu} \rightarrow Q^{\mu\nu} + Y^{\mu\nu} + M^{\mu}\xi^{\nu} - M^{\nu} \xi^{\mu} + D_{\rho} Z^{\mu\nu\rho} \, .
\end{equation}
One can straightforwardly see that the ambiguities in $\Theta^{\mu}$ (eq.\eqref{eq:combthetaamb}) and the ambiguities of $Q^{\mu\nu}$ (eq.\eqref{eq:combambq}) cancel out in the combination of $E_{vv}$ given by
\begin{equation}
    2 E^{\mu\nu} \xi_{\mu} \xi_{\nu} + \xi^{\mu}G_{\mu}(A_{\nu}\xi^{\nu}+\Lambda) = -\xi_{\mu}\Theta^{\mu} + \xi_{\mu} D_{\nu} Q^{\mu\nu} \, .
\end{equation}
This is expected because equations of motion for a theory are unambiguous, so the ambiguities should be canceled out in this combination. The ambiguities, however directly contribute to $\Theta^r$ and $Q^{r\nu}$ and thus, we expect that they should contribute to the entropy current even though they do not contribute to $E_{vv}$. We will see that these ambiguities will contribute in the form of eq.\eqref{eq:addambs}. Thus, the cancellation of ambiguities is more illuminating in our gauge eq.\eqref{eq:metric1}. 

\medskip

We will note down the following results from Appendix \ref{ap:basicsetup} (eq.\eqref{eq:thetaqfinal} and eq.\eqref{eq:evvapfinal}) for the sake of clarity:
\begin{equation}\label{eq:thetaqdefamb}
    \begin{split}
        \Theta^r|_{r=0} &= (1+v \partial_v) \mathcal{A}_{(1)} + v \partial^2_v \mathcal{B}_{(0)} \, , \\
        Q^{r\mu} &= \mathcal{Q}^{r\mu} + v W^{r\mu}_v \, .
    \end{split}
\end{equation}
Using eq.\eqref{eq:mainevvrel} (or eq.\eqref{eq:mainrel}) given by
\begin{equation}\label{eq:mainambrel}
    2 v \, E_{vv}|_{r=0} = \left( - \Theta^r + D_{\mu} Q^{r\mu}\right)|_{r=0} \, ,
\end{equation}
the components of the entropy current can be worked out to be
\begin{equation}\label{eq:entropycurrentcomp}
    \mathcal{J}^v = -\dfrac{1}{2} ( \mathcal{Q}^{rv} + \mathcal{B}_{(0)}) \, , ~~~~~ \mathcal{J}^i = -\dfrac{1}{2} (\mathcal{Q}^{ri} - J^i_{(1)}) \, ,
\end{equation}
where $W^{ri}_v = \partial_v J^i_{(1)} + \mathcal{O}(\epsilon^2)$. We will now analyse the $M-$type (eq.\eqref{eq:lamb}), $Y-$type (eq.\eqref{eq:ytypeamb}) and $Z-$type (eq.\eqref{eq:ztypeamb}) ambiguities in the following. Such ambiguities have been analysed in \cite{Bueno:2016gnv,Speranza:2017gxd,Haehl:2017sot}. They also play a role in the holographic entanglement entropy functionals \cite{Dong:2013qoa,Camps:2013zua,Miao:2014nxa}. We will explicitly show that they are of the form given in eq.\eqref{eq:ambincurrent}.

\subsection*{Analysis of the $M$-type ambiguity:}
 The $M-$type ambiguity in $\Theta^{\mu}$ is given by eq.\eqref{eq:mtypeamb}:
 \begin{equation}
     \Theta^{\mu}[\mathcal{L}_{\xi}g_{\alpha\beta},\mathcal{L}_{\xi}A_{\gamma}+ D_{\gamma}\Lambda] \rightarrow \Theta^{\mu} + \xi^{\mu} D_{\rho} M^{\rho} + D_{\nu} \left[ M^{\mu}\xi^{\nu} - M^{\nu} \xi^{\mu} \right] \, .
 \end{equation}
This is valid only for a diffeomorphism covariant $M^{\mu}$. Thus, at $r=0$, we have
\begin{equation}
    \begin{split}
        \Theta^{r} &\rightarrow \Theta^{r} + D_{\nu} (M^r \xi^{\nu}) \\
        & = \Theta^r + \dfrac{1}{\sqrt{h}}\partial_v \left( \sqrt{h} \, M^r v \right) = \Theta^r + (1+ v \partial_v) M^r + \mathcal{O}(\epsilon^2)\, .
    \end{split}
\end{equation}
Thus, the entropy current of eq.\eqref{eq:entropycurrentcomp} is unaffected by this ambiguity because $\mathcal{B}_{(0)}$ is unchanged. Hence, $M-$type ambiguity does not contribute to the entropy current.

\subsection*{Analysis of the $Y$-type ambiguity:}

We note that this $Y-$type ambiguity has been looked at in the context of defining entropy before in \cite{Bueno:2016gnv,Haehl:2017sot}. Here $Y^{\mu\nu}[\delta g_{\alpha\beta},\delta A_{\gamma}]$ of eq.\eqref{eq:ytypeamb} is assumed to be diffeomorphism covariant and $U(1)$ gauge invariant. Consider $Y^{ri} = Y^{ri}[\mathcal{L}_{\xi}g_{\alpha\beta},\mathcal{L}_{\xi}A_{\gamma}]$. This has the same structure as $\Theta^r$ and thus, one can essentially borrow the analysis of \cite{Bhattacharyya:2021jhr,Biswas:2022grc} to arrive at \footnote{Notation: Here the subscripts are used to denote different terms. It shouldn't be confused with boost weights written within $()$ in subscripts.}
\begin{equation}\label{eq:yristruc}
    \begin{split}
        Y^{ri} &= (1+v \partial_v) \mathcal{Y}^{\prime \, i}_{(1)} + v \partial^2_v \mathcal{Y}^i_{(0)} \, , \\
        \implies \nabla_i Y^{ri} &= (1 + v \partial_v) \nabla_i \mathcal{Y}^{\prime \, i}_{(1)} + v \partial^2_v \nabla_i \mathcal{Y}^i_{(0)} \, .
    \end{split}
\end{equation}
See section 3.2.2 of \cite{Bhattacharyya:2021jhr} and section 6.3 (also eq.(6.32)) of \cite{Biswas:2022grc} for details of this construction. The analysis of section 3.2.2 of \cite{Bhattacharyya:2021jhr} deals with $\delta g_{\alpha\beta}$ terms in $\Theta^r$ and section 6.3 of \cite{Biswas:2022grc} deals with the $\delta A_{\mu}$ terms.

Consider $Y^{rv} = Y^{rv}[\mathcal{L}_{\xi}g_{\alpha\beta},\mathcal{L}_{\xi}A_{\gamma} + D_{\gamma} \Lambda]$. $Y^{rv}$ is boost weight zero and it is linear in $\xi$. It also depends on $\mathcal{L}_{\xi}A_{\gamma}+D_{\gamma}\Lambda$ through $\delta A_{\mu}$. From the analysis of the $\delta A_{\mu}$ terms in eq.(6.26) of \cite{Biswas:2022grc} ((i.e.) the $\delta A_{\mu}$ terms in boost weight zero quantities), the structure of $Y^{rv}$ is of the following form
\begin{equation}\label{eq:yrvstruc}
    Y^{rv} = \Bar{Y}^{rv}_1 + v \mathcal{Y}^{rv}_v + \Bar{Y}^v_2 \partial_v (v A_v + \Lambda) \, .
\end{equation}
There is an additional $v A_v + \Lambda$ piece compared to the structure of $Q^{rv}$ in eq.\eqref{eq:thetaqdefamb}. Here it should be noted that $(v A_{v} + \Lambda) \sim \mathcal{O}(\epsilon)$ according to the analysis presented in Appendix B of \cite{Biswas:2022grc}. Thus, from eq.\eqref{eq:yrvstruc}, we have
\begin{equation}\label{eq:delvyrv}
    \dfrac{1}{\sqrt{h}} \partial_v \left( \sqrt{h}\, Y^{rv}\right) = \dfrac{1}{\sqrt{h}}\partial_v \left( \sqrt{h} \, \Bar{Y}^{rv}_1 \right) + (1 + v \partial_v) \mathcal{Y}^{rv}_v + \dfrac{1}{\sqrt{h}} \partial_v \left( \sqrt{h} \, \Bar{Y}^v_2 \partial_v (v A_v + \Lambda) \right) \, .
\end{equation}
We now have,
\begin{equation}\label{eq:divy}
    \begin{split}
        D_{\nu} Y^{r \nu} &= \dfrac{1}{\sqrt{h}} \partial_v \left( \sqrt{h} Y^{rv} \right) + \nabla_i Y^{ri} \\
        &= (1+v\partial_v) \left[ \mathcal{Y}^{rv}_v + \nabla_i \mathcal{Y}^{\prime \, i}_{(1)} + \dfrac{1}{\sqrt{h}}\partial_v \left( \sqrt{h} \, \Bar{Y}^{rv}_1 \right) \right] \\
        &~~~ + v\partial_v \left[ \dfrac{1}{\sqrt{h}} \partial_v \left[ \sqrt{h} (\nabla_i \mathcal{Y}^i_{(0)} - \Bar{Y}^{rv}_1) \right] \right] + \dfrac{1}{\sqrt{h}} \partial_v \left[ \sqrt{h} \left( \bar{Y}^v_2 (v A_v + \Lambda) \right) \right] + \mathcal{O}(\epsilon^2) \, .
    \end{split}
\end{equation}
In the second step, we have used eq.\eqref{eq:delvyrv}.

eq.\eqref{eq:divy} represents the ambiguity in $\Theta^r$ of the form eq.\eqref{eq:ytypeamb}. Thus, the terms in eq.\eqref{eq:divy} are ambiguities in the definition of $\mathcal{A}_{(1)}$ and $\mathcal{B}_{(0)}$ of $\Theta^r$ of eq.\eqref{eq:thetaqdefamb}. For instance, $\Theta^r \rightarrow \Theta^r + D_{\nu} Y^{r\nu}$ from eq.\eqref{eq:ytypeamb} implies
\begin{equation}\label{eq:a1b0amb}
    \begin{split}
        \mathcal{A}_{(1)} &\rightarrow \mathcal{A}_{(1)} + \mathcal{Y}^{rv}_v + \nabla_i \mathcal{Y}^{\prime \, i}_{(1)} + \dfrac{1}{\sqrt{h}}\partial_v \left( \sqrt{h} \, \Bar{Y}^{rv}_1 \right) \, ,\\
        \mathcal{B}_{(0)} &\rightarrow \mathcal{B}_{(0)} + \nabla_i \mathcal{Y}^i_{(0)} - \bar{Y}^{rv}_1 \, .
    \end{split}
\end{equation}
Now since $Q^{r\nu} \rightarrow Q^{r\nu} + Y^{r\nu}$, the last term proportional to $v A_v + \Lambda$ in eq.\eqref{eq:divy} cancels out in the combination of eq.\eqref{eq:mainambrel}. As $Q^{rv} = \mathcal{Q}^{rv} + v W^{rv}_v$ from eq.\eqref{eq:thetaqdefamb}, $Y^{rv}$ of eq.\eqref{eq:yrvstruc} implies
\begin{equation}\label{eq:qrvyamb}
    \mathcal{Q}^{rv} \rightarrow \mathcal{Q}^{rv} + \bar{Y}^{rv}_1 \, .
\end{equation}
Also, $Q^{ri} = \mathcal{Q}^{ri} + v W^{ri}_v$ from eq.\eqref{eq:thetaqdefamb}, $Y^{ri}$ of eq.\eqref{eq:yristruc} implies
\begin{equation}\label{eq:wrivamb}
    \begin{split}
        \mathcal{Q}^{ri} &\rightarrow \mathcal{Q}^{ri} + \mathcal{Y}^i_{(0)} \, ,\\
        W^{ri}_v &\rightarrow W^{ri}_v + \partial_v \mathcal{Y}^{\prime \, i}_{(1)} + \partial^2_v \mathcal{Y}^i_{(0)} \, .
    \end{split}
\end{equation}
But $W^{ri}_v = \partial_v J^i_{(1)}$, and thus eq.\eqref{eq:wrivamb} implies
\begin{equation}\label{eq:ji1yamb}
     J^i_{(1)} \rightarrow J^i_{(1)} + \mathcal{Y}^{\prime \, i}_{(1)} + \partial_v \mathcal{Y}^i_{(0)} \, .
\end{equation}
Thus collecting eq.\eqref{eq:a1b0amb}, eq.\eqref{eq:qrvyamb}, eq.\eqref{eq:wrivamb} and eq.\eqref{eq:ji1yamb} in eq.\eqref{eq:entropycurrentcomp}, we get
\begin{equation}\label{eq:currentytype}
    \begin{split}
        \mathcal{J}^v &\rightarrow \mathcal{J}^v - \dfrac{1}{2} \nabla_i \mathcal{Y}^i_{(0)} \, , \\
        \mathcal{J}^i &\rightarrow  \mathcal{J}^i + \dfrac{1}{2}\partial_v \mathcal{Y}^i_{(0)} \, .
    \end{split}
\end{equation}
This comprises the effect of $Y-$type ambiguity. $\mathcal{Y}^i_{(0)}$ by definition is $\mathcal{O}(\epsilon)$ and thus this analysis answers why JKM ambiguties arise from $Y-$type ambiguities \cite{Bueno:2016gnv,Haehl:2017sot}. The ambiguities in eq.\eqref{eq:currentytype} are clearly of the form eq.\eqref{eq:ambincurrent}.

\subsection*{Analysis of $Z$-type ambiguity:}
We now analyze the $Z-$type ambiguity of eq.\eqref{eq:ztypeamb}. We have
\begin{equation}
    Q^{rv} \rightarrow Q^{rv} + D_{\rho} Z^{rv\rho} = Q^{rv} + \nabla_i Z^{rvi} \, .
\end{equation}
Thus eq.\eqref{eq:thetaqdefamb} implies
\begin{equation}\label{eq:qrvzamb}
    \mathcal{Q}^{rv} \rightarrow \mathcal{Q}^{rv} + \nabla_i Z^{rvi} \, .
\end{equation}
Similarly
\begin{equation}
    Q^{ri} \rightarrow Q^{ri} + D_{\rho} Z^{ri\rho} = Q^{ri} + \dfrac{1}{\sqrt{h}}\partial_v \left( \sqrt{h}\, Z^{riv} \right) + \nabla_j Z^{rij} \, .
\end{equation}
Notice how $Z^{\mu\nu\rho} = Z^{[\mu\nu\rho]}$ played a crucial role in arriving at this structure. We thus have
\begin{equation}\label{eq:qrizamb}
    \mathcal{Q}^{ri} \rightarrow \mathcal{Q}^{ri} + \dfrac{1}{\sqrt{h}}\partial_v \left( \sqrt{h}\, Z^{riv} \right) + \nabla_j Z^{rij} \, .
\end{equation}
Collecting eq.\eqref{eq:qrvzamb} and eq.\eqref{eq:qrizamb} in eq.\eqref{eq:entropycurrentcomp}, we have
\begin{equation}\label{eq:currentztype}
    \begin{split}
        \mathcal{J}^v &\rightarrow \mathcal{J}^v - \dfrac{1}{2} \nabla_i Z^{rvi} \, , \\
        \mathcal{J}^i &\rightarrow \mathcal{J}^i + \dfrac{1}{2}\dfrac{1}{\sqrt{h}}\partial_v \left( \sqrt{h}\, Z^{rvi} \right) - \dfrac{1}{2}\nabla_j Z^{rij} \, .
    \end{split}
\end{equation}

\subsection*{Summary:}

To summarise, there are three types of Iyer-Wald ambiguities:
\begin{equation}
    \begin{split}
        &\text{(i)} ~~ \mathcal{L} \rightarrow \mathcal{L} + D_{\rho} M^{\rho} \, , \\
        &\text{(ii)} ~~ \Theta^{\mu} \rightarrow \Theta^{\mu} + D_{\nu} Y^{\mu\nu} \, , \\
        &\text{(iii)} ~~ Q^{\mu\nu} \rightarrow Q^{\mu\nu} + D_{\rho} Z^{\mu\nu\rho} \, .
    \end{split}
\end{equation}
The corresponding contributions to the entropy current are given by eq.\eqref{eq:currentytype} and eq.\eqref{eq:currentztype}
\begin{equation}\label{eq:finalambexpr}
    \begin{split}
        &\text{(i)} ~~ \text{No contribution to $\mathcal{J}^v$ and $\mathcal{J}^i$.} \\
        &\text{(ii)} ~~ \mathcal{J}^v \rightarrow \mathcal{J}^v - \dfrac{1}{2}\nabla_i \mathcal{Y}^i_{(0)} \, , ~~ \mathcal{J}^i \rightarrow \mathcal{J}^i + \dfrac{1}{2}\partial_v \mathcal{Y}^i_{(0)} \, , ~~ \mathcal{Y}^i_{(0)} \sim \mathcal{O}(\epsilon) \, . \\
        &\text{(iii)} ~~ \mathcal{J}^v \rightarrow \mathcal{J}^v - \dfrac{1}{2} \nabla_i Z^{rvi} \, , ~~\mathcal{J}^i \rightarrow \mathcal{J}^i + \dfrac{1}{2}\dfrac{1}{\sqrt{h}}\partial_v \left( \sqrt{h}\, Z^{rvi} \right) - \dfrac{1}{2}\nabla_j Z^{rij} \, .
    \end{split}
\end{equation}
Thus, all the entropy current constructions of the previous works \cite{Bhattacharya:2019qal,Bhattacharyya:2021jhr,Biswas:2022grc,Deo:2023vvb} inherently suffer from ambiguities that arise from the covariant phase formalism used \footnote{It is useful to note that the ambiguities (ii) and (iii) do not affect the total entropy which is given by the integral of $\mathcal{J}^v$ over a co-dimension 2 slice of the horizon. This is because the ambiguous terms are total derivatives. This was also recently noted and carefully studied in \cite{Wall:2024lbd}.}. We have proved the second result of our paper in eq.\eqref{eq:ambresult}. The ambiguities in eq.\eqref{eq:ambinevv} are of the form eq.\eqref{eq:ambresult}, and they arise from the ambiguities in $\Theta^r$ and $Q^{r\mu}$.

\medskip

\subsection{Ambiguities in the physical process version of the first law}
\label{ssec:ambppfl}

In this section, we will analyse the ambiguities in the setup of the physical process version of the first law (PPFL) discussed in \cite{Bhattacharyya:2021jhr}. The setup of the second law proved through the null projected EoM structure of eq.\eqref{eq:keyeqn} can be straightforwardly used to argue for a physical process version of the first law \cite{Gao:2001ut,Amsel:2007mh,Chatterjee:2011wj,Bhattacharjee:2014eea,Mishra:2017sqs}. PPFL is a relation between the change of entropy of a Black hole $\delta S$ and the infalling matter generated by a small matter stress tensor $\delta T_{\mu\nu}$, which is given by eq.5.1 of \cite{Bhattacharyya:2021jhr}:
\begin{equation}\label{eq:ppfl1}
    \delta S = 2 \pi \int_{\mathcal{H}} dv \, d^{d-2}x \, \sqrt{h} \chi^{\mu} \xi^{\nu} \delta T_{\mu\nu} \, .
\end{equation}
Section 2.2.3 of \cite{Bhattacharya:2019qal} discusses this equation in detail, and the derivation is given in section 2 of \cite{Jacobson:1995uq}. Here $\mathcal{H}$ is the codimension 1 horizon $r=0$ of eq.\eqref{eq:eqmsol}, $\chi^{\mu} = (\partial_v)^{\mu}$ is the affinely parameterized null generator of the horizon of eq.\eqref{eq:eqmsol} and $\xi^{\mu}$ is the Killing vector eq.\eqref{eq:killing}. We will consider the PPFL between arbitrary time slices of the horizon which naturally leads to an expression for a dynamical entropy for an arbitrary non-stationary slice of the horizon. We will then focus on the effect of ambiguities in this version of PPFL.

\subsubsection*{PPFL between arbitrary time slices of the horizon}
\label{sssec:ppfl}

In writing eq.\eqref{eq:ppfl1}, we made the following assumptions: the horizon is stable under perturbations; perturbations do not shift the location of the horizon \footnote{For example, if in some coordinate system, the dynamical horizon is located at $R = R_H(v)$, then to transform to the coordinates as in \eqref{eq:metric1}, the corresponding transformation would be $r = R - R_H(v)$ \cite{Bhattacharyya:2022nqa}}. The first assumption is a notion of the Weak cosmic censorship hypothesis, which is a desirable property. The second assumption follows because we are working with the Gaussian null coordinate system of eq.\eqref{eq:metric1}. The perturbations are of the form eq.\eqref{eq:pertubs}, and they clearly do not shift the horizon at $r=0$ as the effect of these perturbations is already taken into account in the definition of the $r$-coordinate. An important consequence of this assumption is that the event horizon of the dynamical spacetime coincides with the event horizon of the equilibrium spacetime. This is a well-motivated assumption because we look at situations when a black hole is ``settling down to equilibrium''. This allows us to do the integral over the affine coordinate $v$ on a dynamical horizon because $v$ remains null throughout the entire evolution. As is usually done for the first law, we assume the statement of the zeroth law \cite{IstvánRácz_1996,Bhattacharyya:2022nqa,Davies:2024fut} to be valid.\footnote{On the horizon, the surface gravity $\kappa$ arises from the relation $\xi^{\mu} = \kappa v \chi^{\mu}$.}

\medskip

The RHS of eq.\eqref{eq:ppfl1} represents the mass $\delta M$ falling into the horizon $\mathcal{H}$ and thus the proof of the PPFL will be complete if we can extract a quantity $S$ that satisfies $T\delta S = \delta M$. We will precisely find an $S$ that satisfies the required condition. For this, we first use the on-shell EoM eq.\eqref{eq:eommunu} given by $E_{\mu\nu} = -\delta T_{\mu\nu}$ in eq.\eqref{eq:ppfl1}. After substituting for $\chi^{\mu}$ and $\xi^{\mu}$, we get
\begin{equation}\label{eq:ppfl2}
    \delta S = -2 \pi \int_{\mathcal{H}} dv \, d^{d-2}x \, \sqrt{h} \, v E_{vv} \, .
\end{equation}
Using eq.\eqref{eq:keyeqn} in eq.\eqref{eq:ppfl2}, section 5 of \cite{Bhattacharyya:2021jhr} extracted an $S$ from eq.\eqref{eq:ppfl2} thereby proving the PPFL. In order to work this out, we must integrate over $v$ from some past time slice to a future time slice of codimension 1 horizon. Thus, we have
\begin{equation}\label{eq:ppfl3}
    \delta S = -2 \pi \int^{v=v_1}_{v=v_2} dv \, \int_{\mathcal{H}_v} d^{d-2}x \, \sqrt{h} \, v E_{vv} \, .
\end{equation}
Here $\mathcal{H}_v$ denotes the constant $v$ spatial slices. In \cite{Bhattacharyya:2021jhr}, the authors considered the case of entropy increase from an initial equilibrium configuration in the far past to a final equilibrium configuration in the far future. Thus, the entropy increase is given by the difference of the equilibrium Wald entropy evaluated in both slices. This is a geometric quantity that is clearly gauge invariant. \cite{Bhattacharyya:2022njk} argued for eq.\eqref{eq:jvjitransf} as a consequence of gauge invariance of this difference of Wald entropy. Thus, we naturally have a notion of gauge invariance of entropy production.

\medskip

However, there is nothing special about the choice of constant-$v$ slices made in 
\cite{Bhattacharyya:2021jhr}. The setup of section 5 of \cite{Bhattacharyya:2021jhr} can be used to argue for a PPFL for arbitrary $v_1$ and $v_2$. Such an analysis was already carried out for Lovelock theories in \cite{Mishra:2017sqs}. As a byproduct of this analysis, we can obtain an expression for the entropy of a dynamical black hole in an arbitrary diffeomorphism invariant theory of gravity for arbitrary time slices on the horizon. Such an expression would be consistent with the PPFL. To derive that expression, we first substitute eq.\eqref{eq:keyeqn} in eq.\eqref{eq:ppfl3} to obtain
\begin{equation}\label{eq:ppfl4}
    \delta S = - \int^{v=v_1}_{v=v_2} dv \, \int_{\mathcal{H}_v} d^{d-2}x \, \sqrt{h} \, v \partial_v \Theta \, ,
\end{equation}
where $\Theta$ is the generalized expansion given by
\begin{equation}\label{eq:gentheta}
    \Theta = \dfrac{1}{\sqrt{h}} \partial_v \left( \sqrt{h} \, \mathcal{J}^v \right) + \nabla_i \mathcal{J}^i \, .
\end{equation}
In writing eq.\eqref{eq:ppfl4}, we have normalized the Lagrangian to get rid of the $2\pi$ factor. We can now basically follow the steps of section 5 of \cite{Bhattacharyya:2021jhr}. One can do an integration by parts on eq.\eqref{eq:ppfl4} to obtain
\begin{equation}\label{eq:ppfl5}
    \begin{split}
        \delta S &= - \left[ \int_{\mathcal{H}_v} d^{d-2}x \, \sqrt{h} \, v \, \Theta \right]^{v=v_1}_{v=v_2} + \int^{v=v_1}_{v=v_2} dv \int_{\mathcal{H}_v} d^{d-2}x \, \partial_v (v \sqrt{h}) \, \Theta \\
        &= - \left[ \int_{\mathcal{H}_v} d^{d-2}x \, \sqrt{h} \, v \, \Theta \right]^{v=v_1}_{v=v_2} + \int^{v=v_1}_{v=v_2} dv \int_{\mathcal{H}_v} d^{d-2}x \, \sqrt{h} \, \Theta + \mathcal{O}(\epsilon^2) \, ,
    \end{split}
\end{equation}
where the second step arises from the fact that $\Theta \sim \mathcal{O}(\epsilon)$ and $\partial_v (v \sqrt{h}) = \sqrt{h} + \mathcal{O}(\epsilon)$. Now,
\begin{equation}\label{eq:thetaint}
   \begin{split}
        \int^{v=v_1}_{v=v_2} dv\int_{\mathcal{H}_v} d^{d-2}x \sqrt{h} \, \Theta &= \int^{v=v_1}_{v=v_2} dv\int_{\mathcal{H}_v} d^{d-2}x \sqrt{h} \, \left[ \dfrac{1}{\sqrt{h}} \partial_v \left( \sqrt{h} \, \mathcal{J}^v \right) + \nabla_i \mathcal{J}^i \right] \\
        &= \int^{v=v_1}_{v=v_2} dv\int_{\mathcal{H}_v} d^{d-2}x \partial_v \left( \sqrt{h} \mathcal{J}^v \right) \\
         &= \left[ \int d^{d-2}x \, \sqrt{h} \, \mathcal{J}^v \right]^{v=v_1}_{v=v_2}
   \end{split}
\end{equation}
In the second step, we have assumed that the constant $v$ slices $\mathcal{H}_v$ are compact to integrate out $\nabla_i \mathcal{J}^i$. Substituting the result of eq.\eqref{eq:thetaint} in eq.\eqref{eq:ppfl5}, we get
\begin{equation}\label{eq:ppflfinal}
    \delta S = -\left[ \int_{\mathcal{H}_v} d^{d-2}x \, \sqrt{h} \, v \, \Theta \right]^{v=v_1}_{v=v_2} + \left[ \int_{\mathcal{H}_v} d^{d-2}x \, \sqrt{h} \, \mathcal{J}^v \right]^{v=v_1}_{v=v_2} \, .
\end{equation}
eq.\eqref{eq:ppflfinal} gives an expression for the entropy of a dynamical black hole for an arbitrary time slice of the horizon in a general diffeomorphism invariant theory of gravity. This is given by
\begin{equation}\label{eq:entropydyn}
    S|_{\mathcal{H}_v} = \int_{\mathcal{H}_v} d^{d-2}x \, \sqrt{h} \, \mathcal{J}^v - \int_{\mathcal{H}_v} d^{d-2}x \, \sqrt{h} \, v \, \Theta
\end{equation}
At this point, it is useful to mention that the presence of an additional boundary term proportional to $v\,\Theta$ in eq.\eqref{eq:ppflfinal} for arbitrary time slices of the horizon has already been emphasized in \cite{Mishra:2017sqs}. By changing the null parameterization to a non-affine parameterization, the authors of \cite{Mishra:2017sqs} have interpreted this extra term as a change in energy $\delta E$ of the horizon membrane obtained through the membrane paradigm picture.

\medskip

If we consider, Einstein gravity given by $L=R$, we have $\mathcal{J}^v = 1, \, \mathcal{J}^i =0$. Thus, eq.\eqref{eq:entropydyn} becomes
\begin{equation}\label{eq:seinstein}
   \begin{split}
        S[L=R]|_{\mathcal{H}_v} &= \int_{\mathcal{H}_v} d^{d-2}x \, \sqrt{h} - \int_{\mathcal{H}_v} d^{d-2}x \, \sqrt{h} \, v \, \theta \\
        &= A[\mathcal{H}_v] - \int_{\mathcal{H}_v} d^{d-2}x \, \sqrt{h} \, v \, \theta \, ,
   \end{split}
\end{equation}
where $A[\mathcal{H}_v]$ \footnote{In our choice of normalization, we dropped the $2\pi$ factor in eq.\eqref{eq:ppfl4} and thus we don't get the standard $\frac{1}{4}$ factor. If the Lagrangian is of the form $\frac{1}{16\pi}R$, then we do get the usual $\frac{1}{4}$ factor from the Wald entropy.} denotes the area of the constant $v$ slice $\mathcal{H}_v$ and $\theta = \dfrac{1}{\sqrt{h}}\partial_v \sqrt{h}$ is the expansion of the null generators $\chi^{\mu} = (\partial_v)^{\mu}$ \footnote{$S$ of eq.\eqref{eq:seinstein} was recently obtained in \cite{Hollands:2024vbe} using covariant phase methods of \cite{PhysRevD.61.084027}. This $S$ was interpreted to be the Area of the apparent horizon.}. For arbitrary higher derivative theories of gravity of the form $\mathcal{L}$ of eq.\eqref{eq:lag1}, the dynamical entropy is given by eq.\eqref{eq:entropydyn}. The $\Theta$ clearly contains the information of the entropy current $\mathcal{J}^i$ through eq.\eqref{eq:gentheta}. We could alternately choose to integrate out $\mathcal{J}^i$ by assuming $\mathcal{H}_v$ is compact to obtain \footnote{eq.\eqref{eq:waldentdyn} was obtained in \cite{Hollands:2024vbe}. The entropy expression also satisfies the PPFL between arbitrary time slices.}
\begin{equation}\label{eq:waldentdyn}
    S|_{\mathcal{H}_v} = \int_{\mathcal{H}_v} d^{d-2}x \, \sqrt{h} \, \mathcal{J}^v - v \partial_v \left[ \int_{\mathcal{H}_v} d^{d-2}x \, \sqrt{h} \mathcal{J}^v\right] \, .
\end{equation}
It is useful to note that the expressions in eq.\eqref{eq:entropydyn} and eq.\eqref{eq:waldentdyn} are valid for arbitrary non-minimally coupled theories of the form eq.\eqref{eq:lag1}. This is because these quantities are directly derived from the structure of $E_{vv}$ of the eq.\eqref{eq:keyeqn}, which in turn follows from the fact that we have a Killing horizon at $r=0$ for an equilibrium configuration. The structure of $E_{vv}$ for eq.\eqref{eq:lag1} has already been established in \cite{Biswas:2022grc} and it takes the form of eq.\eqref{eq:keyeqn}. The currents $\mathcal{J}^v$ and $\mathcal{J}^i$ would receive contributions from non-minimal matter couplings. A linearized PPFL for such couplings has been proved in section 7 of \cite{Biswas:2022grc}. The subtlety of carefully incorporating both the matter and the gravitational contributions from non-minimal couplings to the PPFL has been cleanly worked out in section 7 of \cite{Biswas:2022grc}. 

\medskip

We now study the effect of the Iyer-Wald ambiguities summarized in eq.\eqref{eq:finalambexpr} on the general expression for the dynamical entropy given by eq.\eqref{eq:entropydyn}. It is clear from eq.\eqref{eq:gentheta} that $\Theta$ remains invariant under the ambiguities of eq.\eqref{eq:finalambexpr}. We should only analyse the effect of ambiguities on by $\mathcal{J}^v$, which is given by $\mathcal{J}^v \to \mathcal{J}^v + \nabla_i A^i$. This is a total derivative term that drops out when you integrate over a compact $\mathcal{H}_v$. Thus, the Iyer-Wald ambiguities do not affect the PPFL up to linear order in perturbations. This has also been emphasized recently in \cite{Wall:2024lbd}.

\section{Verification of the proof}
\label{sec:verproof}

In this section, we will consider explicit examples of Lagrangians considered in \cite{Bhattacharyya:2021jhr} and \cite{Biswas:2022grc} to illustrate the general formulae for the transformation of the entropy currents derived in eq.\eqref{eq:jvjifinaltransf}. In what follows, we will do a brute force computation of how the components of entropy current transform under a reparametrization eq.\eqref{eq:repara}. We will extensively use the transformation properties of various structures collected in Appendix \ref{ap:reparadetails} to analyse how the components of the entropy current in explicit examples transform. In all cases, the brute force results match the prediction eq.\eqref{eq:jvjifinaltransf} of the abstract proof.

\subsection{Non-minimally coupled scalar field}

We first consider a case of Non-minimally coupled scalar field with a Lagrangian given by
\begin{equation}\label{sf}
    \mathcal{L}{(\phi, g_{\mu\nu})} = R_{\mu\nu} (D^\mu \phi)(D^\nu \phi) \, ,
\end{equation}

for which the components of the entropy current were worked out to be \cite{Biswas:2022grc} (eq.(5.26) of \cite{Biswas:2022grc})
\begin{equation}
    \mathcal{J}^v = (\partial_v \phi)(\partial_r \phi) \, , ~~~~~~~~~~ \mathcal{J}^i = (\nabla^i \phi) (\partial_v \phi) \, .
\end{equation}
Under a reparametrization of the form eq.\eqref{eq:repara}, the components of the entropy current transform as
\begin{equation}
    \begin{split}
        \mathcal{J}^v &= (\partial_r \phi)(\partial_v \phi) \\
        &=  (\partial_{\rho} \phi)(\partial_{\tau} \phi) + \tau \xi_a (\partial^a \phi)(\partial_{\tau} \phi) \\
        &= \widetilde{\mathcal{J}}^{\tau} + \tau \xi_a \widetilde{\mathcal{J}}^{a} \, ,\\
    \mathcal{J}^i&=(\nabla^i \phi) (\partial_v \phi)\\
    &= e^{-\zeta} \widetilde{ \mathcal{J}}^i \, .
    \end{split}
\end{equation}
These expressions agree with the general result eq.\eqref{eq:jvjifinaltransf} once we see that
\begin{equation}
    \widetilde{E}^{\rho i \rho j}_R = \mathcal{O}(\epsilon^2) \implies \widetilde{L}^{ij}_{(1)} = \mathcal{O}(\epsilon^2) \, .
\end{equation}

One important point to notice here is that $\mathcal{B}_{(0)}=\mathcal{O}(\epsilon^2)$ remains $\mathcal{O}(\epsilon^2)$ under the transformation given by eq.\eqref{eq:botransf}. This can be seen from the following transformations as 
\begin{equation}
    \begin{split}
        \mathcal{Q}^{rv}&=-2 (\partial_v \phi)(\partial_r \phi)=\widetilde{\mathcal{Q}}^{\rho\tau} + 2\tau\xi_{a}\widetilde{\mathcal{Q}}^{\rho a} \, ,\\
        \mathcal{Q}^{ri}&= - (\nabla^i \phi) (\partial_v \phi)= e^{-\zeta} \mathcal{Q}^{\rho i} \, , \\
       J^i_{(1)}&= (\nabla^i \phi) (\partial_v \phi)= e^{-\zeta} \widetilde{J}^i_{(1)} \, .
    \end{split}
\end{equation}
The answers for these structures have been worked out in eq.(5.25) of \cite{Biswas:2022grc}. Therefore, under the reparametrization, $\mathcal{B}_{(0)}$ transforms to
\begin{equation}
    \mathcal{B}_{(0)} = \widetilde{\mathcal{B}}_{(0)} - \tau \xi_i \widetilde{\mathcal{Q}}^{\rho i} - \tau \xi_i \widetilde{J}^i_{(1)} - \tau \widetilde{\nabla}_i \left( 4 \xi_j \widetilde {L}^{ij}_{(1)}\right) + 4 \tau \xi_i \xi_j \widetilde{L}^{ij}_{(1)} = \mathcal{O}(\epsilon^2) \, .
\end{equation}
This is an important validation of the general result in eq.\eqref{eq:jvjifinaltransf} because a simple reparametrization shouldn't generate any non-trivial expression from trivial expressions.

\subsection{Non-minimally coupled gauge field}

Consider a non-minimally coupled gauge field with a Lagrangian of the form
\begin{equation}\label{gf}
    \mathcal{L}{(A_{\mu}, g_{\mu\nu})}= R_{\mu\nu\rho\sigma} F^{\mu\nu}F^{\rho\sigma} \, .
\end{equation}
The components of the entropy current were worked out to be \cite{Biswas:2022grc} (eq.(6.44) of \cite{Biswas:2022grc})
\begin{equation}
    \mathcal{J}^v = -2 F^{rv} F^{rv} \, , ~~~~~~ \mathcal{J}^i = -4 F^{rv} F^{ri} \, . 
\end{equation}
Since $F^{\mu\nu}$ is a covariant tensor, the transformation of $\mathcal{J}^v$ and $\mathcal{J}^i$ under eq.\eqref{eq:repara} can be worked out to be (upto linear order in $\epsilon$)
\begin{equation}
    \begin{split}
        \mathcal{J}^v&= -2  F^{rv}F^{rv} =-2\frac{\partial{r}}{\partial{\rho}}\frac{\partial{v}}{\partial{\tau}}F^{\rho\tau}\frac{\partial{r}}{\partial{\rho}}\frac{\partial{v}}{\partial{\tau}}F^{\rho\tau}  \\ 
     &=\widetilde{\mathcal{J}}^{\tau}-4\tau\xi_a F^{\rho a}F^{\rho\tau} \\
     &= \widetilde{\mathcal{J}}^{\tau} + \tau \xi_a \widetilde{\mathcal{J}}^{a} \, ,\\
    \mathcal{J}^i&=-4 F^{rv}F^{ri}= e^{-\zeta} \widetilde{ \mathcal{J}}^i \, .
    \end{split}
\end{equation}
These expressions agree with eq.\eqref{eq:jvjifinaltransf} once we use the fact that
\begin{equation}
    \widetilde{E}^{\rho i \rho j}_R = \mathcal{O}(\epsilon^2) \implies \widetilde{L}^{ij}_{(1)} = \mathcal{O}(\epsilon^2) \, .
\end{equation}

Similar to the non-minimally coupled scalar case, here also we have $\mathcal{B}_{(0)} = \mathcal{O}(\epsilon^2)$. To see that it remains $\mathcal{O}(\epsilon^2)$ under reparametrization, we use the expressions for $\mathcal{Q}^{rv} $,$\mathcal{Q}^{ri} $ and $J^i_{(1)}$ worked out in eq.(6.47) of \cite{Biswas:2022grc}:
\begin{equation}
    \begin{split}
     \mathcal{Q}^{rv}&=4 F^{rv}F^{rv}  \\
     \mathcal{Q}^{ri}&=4 F^{ri}F^{rv}  \\
     J^i_{(1)}&=-4 F^{ri}F^{rv}
    \end{split}
\end{equation}
Thus, we straightforwardly have
\begin{equation}
    \mathcal{B}_{(0)} = \widetilde{\mathcal{B}}_{(0)} - \tau \xi_i \widetilde{\mathcal{Q}}^{\rho i} - \tau \xi_i \widetilde{J}^i_{(1)} - \tau \widetilde{\nabla}_i \left( 4 \xi_j \widetilde{L}^{ij}_{(1)}\right) + 4 \tau \xi_i \xi_j \widetilde{L}^{ij}_{(1)} = \mathcal{O}(\epsilon^2) \, .
\end{equation}
where, again, we see that the reparametrization of the geodesics doesn't affect these trivial terms.

\subsection{Ricci scalar squared theory}

In the following subsections, we will consider the effects of the reparametrization on some higher-curvature gravity theories. Starting with the Ricci scalar squared theory, the Lagrangian is given in terms of the Ricci scalar for the full metric ($R$) as
 \begin{equation}
     \mathcal{L}= R^2 \, .
 \end{equation}
The expressions for $\mathcal{J}^v$ and $\mathcal{J}^i$ corresponding to this theory were worked out in Table 1 of \cite{Bhattacharya:2019qal} (also see Table 7 above eq.(4.10) of \cite{Bhattacharyya:2021jhr}):
\begin{equation}
    \begin{split}
        \mathcal{J}^v = 2 R \, ,  ~~~~~~
         \mathcal{J}^i = 0 \, .
    \end{split}
\end{equation}
Since $R$ is the Ricci tensor for the full metric, it is invariant under a coordinate transformation. One can also see this if we use the complicated metric transformations of eq.\eqref{eq:metrictransf} in eq.\eqref{eq:ricciriemann}. This serves as a consistency check of the formulae derived in eq.\eqref{eq:metrictransf}. We thus have,
\begin{equation}
    \begin{split}
         \widetilde{\mathcal{J}}^{\tau} = 2 \widetilde{R} \, , ~~~~~~
         \widetilde{\mathcal{J}}^i = 0 \, .
    \end{split}
\end{equation}
These expressions exactly match with the general expression given in eq.\eqref{eq:jvjifinaltransf} if one uses
\begin{equation}
    \widetilde{E}^{\rho i \rho j}_R = 0 \implies \widetilde{L}^{ij}_{(1)} = 0 \, .
\end{equation}

\subsection{Ricci tensor squared theory}

Next, we consider the Ricci tensor squared theory with the Lagrangian of the form
 \begin{equation}
     \mathcal{L}= R_{\mu\nu}R^{\mu\nu} \, .
 \end{equation}
The corresponding $\mathcal{J}^v$ and $\mathcal{J}^i$ were worked out in Table 1 of \cite{Bhattacharya:2019qal} to be:
\begin{equation}
    \begin{split}
    \mathcal{J}^v&=2 R_{r v}-K \bar{K} \, ,\\
    \mathcal{J}^i&=-h^{i j}\partial _v\omega _j+2 \nabla _jK^{i j}-\nabla ^iK \, . \\    
    \end{split}
\end{equation}
In $(\rho,\tau,y^a)$ coordinates, they can be written as
\begin{equation}
    \begin{split}
     \widetilde{\mathcal{J}}^{\tau}&=2 \widetilde{R}_{\rho \tau }-\widetilde{K} \widetilde{\bar{K}} \, , \\  
     \widetilde{\mathcal{J}}^i&=-\widetilde{h}^{i j}\partial _{\tau }\widetilde{\omega }_j+2 \widetilde{\nabla }_j\widetilde{K}^{i j}-\widetilde{\nabla }^i\widetilde{K} \, . \\
    \end{split}
\end{equation}
Using the change of structures  under reparametrization eq.\eqref{eq:metrictransf} and eq.\eqref{eq:basistransf}, the transformation of the components of the entropy current can be obtained as:
\begin{equation}
    \begin{split}
        \mathcal{J}^v-\widetilde{\mathcal{J}}^{\tau}&=\tau \left(2 \xi _i\widetilde{\nabla }_j\widetilde{K}^{i j}-2\xi \cdot \widetilde{\nabla }\widetilde{K}-\widetilde{K}\widetilde{\nabla }\cdot \xi -\xi \cdot \partial _{\tau }\widetilde{\omega }+\xi ^2 \widetilde{K}\right)+\tau ^2\xi ^2\partial _{\tau }\widetilde{K} \, ,\\
        e^{\zeta }\mathcal{J}^i-\widetilde{\mathcal{J}}^i&=\xi _j\widetilde{h}^{i j}\left(1+\tau  \partial _{\tau }\right)\widetilde{K} \, .
    \end{split}
\end{equation}
These expressions match in the linear order with the general expression given in eq.\eqref{eq:jvjifinaltransf} if we use
\begin{equation}
    \widetilde{E}^{\rho i \rho j}_R = \partial_{\tau} \left( -\dfrac{1}{2}\widetilde{h}^{ij}\widetilde{K} \right) \implies \widetilde{L}^{ij}_{(1)} = - \dfrac{1}{2} \widetilde{h}^{ij}\widetilde{K} \, .
\end{equation}

\subsection{Riemann tensor squared theory}

The Lagrangian for the Riemann Tensor Squared Theory is 
\begin{equation}
     \mathcal{L}= R_{\mu\nu\rho\sigma}R^{\mu\nu\rho\sigma} \, .
\end{equation}
The entropy current and density were worked out in Table 1 of \cite{Bhattacharya:2019qal} to be:
\begin{equation}
    \begin{split}
       \mathcal{J}^v&=-4 \left(K^{i j} \bar{K}_{i j}+R_{r v r v}\right)=-4 \left(K^{i j} \bar{K}_{i j}+X +\frac{\omega ^2}{4}\right) \, , \\
        \mathcal{J}^i &= 4\left( \nabla _jK^{i j}-h^{i j}\partial _v\omega _j\right) \, . \\
    \end{split}
\end{equation}
In $(\rho,\tau,y^a)$ coordinates, they can be written as
\begin{equation}
    \begin{split}
       \widetilde{\mathcal{J}}^{\tau}&=-4 \left(\widetilde{K}^{i j} \widetilde{\bar{K}}_{i j}+\widetilde{R}_{\rho \tau \rho \tau }\right)=-4 \left(\widetilde{K}^{i j} \widetilde{\bar{K}}_{i j}+\widetilde{X}+\frac{\widetilde{\omega }^2}{4}\right) \, , \\
        \widetilde{\mathcal{J}}^i&= 4\left(\widetilde{\nabla }_j\widetilde{K}^{i j}-\widetilde{h}^{i j}\partial _{\tau }\widetilde{\omega }_j\right) \, . \\       
    \end{split}
\end{equation}

Again, using the change of structures  under reparametrization eq.\eqref{eq:metrictransf} and eq.\eqref{eq:basistransf}, we get the transformation of the components of the entropy current to be:
\begin{equation}
    \begin{split}
        \mathcal{J}^v-\widetilde{\mathcal{J}}^{\tau}&=4 \tau  \left(\xi _a \xi _b \widetilde{K}^{a b}-\widetilde{K}^{a b}\widetilde{\nabla }_a\xi _b-\xi _a\widetilde{h}^{a b}\partial _{\tau }\widetilde{\omega }_b\right)+4 \tau ^2 \left(\xi _a \xi _b \partial_\tau \widetilde{K}^{a b}\right) \, ,\\
        \mathcal{J}^i-e^{-\zeta } \widetilde{\mathcal{J}}^i&=4e^{-\zeta }\xi _j \left(1+\tau  \partial _{\tau }\right)\widetilde{K}^{i j} \, .
    \end{split}
\end{equation}
These expressions exactly match with the general expression given in eq.\eqref{eq:jvjifinaltransf} if we use 
\begin{equation}
    \widetilde{E}^{\rho i \rho j}_R = - 2\partial_{\tau} \widetilde{K}^{ij} \implies \widetilde{L}^{ij}_{(1)} = - 2 \widetilde{K}^{ij} \, .
\end{equation}

\subsection{Gauss Bonnet theory}

Combining the Lagrangians of the last three subsections, the Gauss-Bonnet Lagrangian can be obtained as
\begin{equation}
     \mathcal{L}= R^2 - 4 R_{\mu\nu}R^{\mu\nu} +  R_{\mu\nu\rho\sigma}R^{\mu\nu\rho\sigma}  \, .
\end{equation}
The transformations of the entropy currents were explicitly worked out in section 4 of \cite{Bhattacharyya:2022njk}. The transformation of the components of the entropy current under reparametrization can be calculated from the previous results to be
\begin{equation}
    \begin{split}
     \mathcal{J}^v-  \widetilde{\mathcal{J}}^{\tau}&= 2\mathcal{R}-2\widetilde{\mathcal{R}}\\&=2 \left[-2\tau  \left( -\xi \cdot \widetilde{K}\cdot \xi  +4  \widetilde{K}^{i j} \widetilde{\nabla} _i\xi _j+4 \xi _i \widetilde{\nabla} _j\widetilde{K}^{i j}\right. \right.\\
     &\left. \left.-2 \xi \cdot \widetilde{\nabla} \widetilde{K} + \xi ^2 \widetilde{K} - 2 \xi _i\widetilde{\nabla _j}\widetilde{K}^{i j}-3 \widetilde{K}^{i j}\widetilde{\nabla _j}\xi _i
     - \widetilde{K} \widetilde{\nabla }\cdot \xi \right) \right. \\
     &\left.- 2 \tau ^2\left( \xi ^2 \partial_\tau \widetilde{K}-\xi \cdot \partial_\tau \widetilde{K}\cdot \xi\right) \right] \, ,\\
     \mathcal{J}^i- e^{-\zeta} \widetilde{ \mathcal{J}}^i&=4 e^{-\zeta}\left(\xi_j \widetilde{K}^{ij}+\tau\xi_j \partial_\tau \widetilde{K}^{ij}- \xi_j \widetilde{h}^{ij} \widetilde{K}- \tau \xi_j \widetilde{h}^{ij} \partial_\tau \widetilde{K}\right) \, ,
    \end{split}
\end{equation}
which match exactly with the ones obtained in \cite{Bhattacharyya:2022njk}. We see that we can obtain the expressions already derived there from the general result of eq.\eqref{eq:jvjifinaltransf} with 
\begin{equation}
    \widetilde{E}^{\rho i \rho j}_R = \partial_{\tau}\left(2\widetilde{h}^{ij}\widetilde{K}-2 \widetilde{K}^{ij}\right) \implies \widetilde{L}^{ij}_{(1)} = 2\widetilde{h}^{ij}\widetilde{K}-2 \widetilde{K}^{ij} \, .
\end{equation}

\medskip

This completes our check of the validation of eq.\eqref{eq:jvjifinaltransf} for explicit example theories.

\section{Gauge covariance to the non-linear order - a case study}
\label{sec:nonlinear}

In this section, we make some preliminary observations of the consequences of a reparametrization of the form eq.\eqref{eq:repara} in the components of the entropy current to the non-linear order in the amplitude of dynamics. The analysis in section \ref{sec:jvjitransfgen} and \ref{sec:ambcurrent} was restricted to the linear order in dynamics. We want to see if eq.\eqref{eq:jvjitransf} continues to hold after including the non-linear terms are added. For this, we require the non-linear terms in eq.\eqref{eq:keyeqn}. Treating higher derivative terms in a Lagrangian like eq.\eqref{eq:lpure} in an Effective field theory (EFT) sense, one can show that \cite{Hollands:2022fkn,Davies:2022xdq,Davies:2023qaa} the structure of $E_{vv}$ is given by
\begin{equation}\label{eq:evvnonlinear}
    E_{vv}|_{r=0} = \underbrace{\partial_v \left( \dfrac{1}{\sqrt{h}} \partial_v \left( \sqrt{h} \, \mathcal{J}^v \right) + \nabla_i \mathcal{J}^i \right)}_{E} \underbrace{- (K_{ij}+X_{ij})(K^{ij}+X^{ij}) + D_i Y^i}_{F} + \mathcal{O}(l^N) 
\end{equation}
This structure was first argued for Lovelock class of theories in \cite{Bhattacharyya:2016xfs} \footnote{This structure was also recently argued for non-minimally coupled Lagrangians of the form eq.\eqref{eq:lag1} \cite{Davies:2024fut} treated in an EFT sense.}. The Lagrangian in eq.\eqref{eq:lpure} will be treated like a formal sum of terms starting with the Einstein two derivative gravity with higher derivatives suppressed by the UV length scale $l$. In the above eq.\eqref{eq:evvnonlinear}, we assume that the EFT Lagrangian is explicitly known up to $N$ or fewer derivatives only.

In eq.\eqref{eq:evvnonlinear}, we have chosen to indicate the entropy current structure as `$E$' and the non-linear structures as `$F$'. We want to see if $E$ and $F$ individually transform like eq.\eqref{eq:evvtrans}. This is the case for the $(2+1)$ dimensional pure gravitational Chern-Simons term worked out in section 4.1 of \cite{Biswas:2022grc}. We want to take the simplest example of the Gauss-Bonnet theory to check whether $E$ and $F$ transform like eq.\eqref{eq:evvtrans}. We will see that there is an additional structure to the transformation, and $E$ and $F$ won't individually transform as eq.\eqref{eq:evvtrans}.

\subsubsection*{Gauss Bonnet theory}
We will start with an effective field approach with Gauss-Bonnet theory as the sub-leading order term in the Lagrangian:
\begin{equation}\label{eq:egblag}
    \mathcal{L}_{\text{gauss}}= \frac{1}{2} \int{d^d x \sqrt{-g} \bigg[R +  l^2 (R^2-4 R_{\mu\nu}R^{\mu\nu}+4 R_{\mu\nu\rho\sigma}R^{\mu\nu\rho\sigma})\bigg]} \, .
\end{equation}
The papers \cite{Bhattacharyya:2016xfs,Hollands:2022fkn} have investigated $E_{vv}$ at the non-linear level. Following \cite{Hollands:2022fkn} and eq.\eqref{eq:evvnonlinear}, $F$ is given by
\begin{equation}\label{eq:nonlinear}
   F = E_{vv}|_{r=0}-\partial_v \left( \dfrac{1}{\sqrt{h}} \partial_v \left( \sqrt{h} \, \mathcal{J}^v \right) + \nabla_i \mathcal{J}^i \right) \, ,
\end{equation}
where $F$ contains quadratic or higher order quantities of positive boost weight up to $\mathcal{O}(\epsilon^2)$ terms as indicated in eq.\eqref{eq:evvnonlinear}. The second term on the RHS is $E$, the local entropy production term which contains both linear and non-linear pieces as denoted below 
\begin{equation}\label{eq:lepbreak}
  \partial_v \left( \dfrac{1}{\sqrt{h}} \partial_v \left( \sqrt{h} \, \mathcal{J}^v \right) + \nabla_i \mathcal{J}^i \right)   = A(\epsilon) + B(\epsilon^2) \,  .
\end{equation}
We have shown that local entropy production is gauge covariant under a reparametrization as in eq.\eqref{eq:jvjitransf}. This split of the linear and non-linear terms will show the additional structure in the transformation properties, as will be clear soon.

\medskip

For the Einstein-Gauss-Bonnet theory of eq.\eqref{eq:egblag}, the entropy density $(J^v)$ and entropy current $(J^i)$ on the horizon are given by Table 7 and eq.(4.14) of \cite{Bhattacharyya:2021jhr}:
\begin{equation}
    \begin{split}
    J^v&= 1+2 l^2 \mathcal{R}\, , \\
    J^i&= l^2\bigl[-4 \nabla_j K^{ij}+4\nabla^i 
    K\bigr] \, .
    \end{split}
\end{equation}
Therefore, the local entropy production term can be expanded up to $\mathcal{O}(\epsilon ^2)$  in terms of metric components as
\begin{equation}{\label{eq:lepGB}}
\begin{split}
\partial _v\left(\frac{1}{\sqrt{h}}\partial _v\left(\sqrt{h} J^v\right)+\nabla _iJ^i\right)&= l^2 [2\mathcal{R}\partial _vK+2K\partial _v\mathcal{R}+2\partial _v{}^2\mathcal{R}+4\partial _v\nabla _i\nabla ^iK\\&-4\partial _v\nabla _i\nabla _jK^{i j}]+\mathcal{O}( \epsilon ^3) \, .
  \end{split}
\end{equation}
Thus, we get the explicit form of $A(\epsilon)$ and $B(\epsilon^2)$ as defined in eq.\eqref{eq:lepbreak}
\begin{equation}
\begin{split}
    A(\epsilon)&= l^2\big[2\mathcal{R} \partial _vK+2\partial _v{}^2\mathcal{R}+4\partial _v\nabla _i\nabla ^iK-4\partial _v\nabla _i\nabla _j K^{i j}\big] \, ,  \\
    B(\epsilon^2) &= l^2\big[2 K \partial_v \mathcal{R} \big] \, . 
\end{split}
\end{equation}

\medskip
 
We now look into the structure of $E_{vv}$ in eq.\eqref{eq:nonlinear} for eq.\eqref{eq:egblag}, which is given in eq.(3.57) of \cite{Bhattacharya:2019qal}
\begin{equation}\label{eq:evv1}
    E_{vv}= R_{vv}+ l^2 (2RR_{vv}-4R^{\alpha\beta}R_{v\alpha v \beta}-4 R^{~\alpha}_v R_{v\alpha} + 2 R^{~\alpha \beta \sigma}_{v}R_{v \alpha \beta \sigma}) \, .
\end{equation}
Eq.\eqref{eq:evv1} can be written explicitly in terms of metric components as 
\begin{equation}{\label{eq:evv2}}
 \begin{split}
 E_{v v}&=(-\partial_v K- K_{ij}K^{ij})+2 l^2 \bigl[ 2\mathcal{R}_{i j}\partial _vK^{i j}+6 \mathcal{R}_{i j} K_k^i K^{j k}-\mathcal{R} \partial _vK-\mathcal{R} K_{i j} K^{i j}\\&-2 \omega ^k K_{j k} \nabla _iK^{i j}+2 \omega _i K^{i j} \nabla _jK
 +K \omega _i \omega _j K^{i j}-2 \nabla _iK^{i j} \nabla _kK_j^k+4 \nabla _iK \nabla _jK^{i j}\\
 &+2 K \omega _i \nabla _jK^{i j}-2\nabla _iK \nabla ^iK-2K \omega _i\nabla ^iK -\frac{1}{2}K^2 \omega ^2 +\frac{1}{2} \omega ^2 K_{i j} K^{i j} -\omega _i \omega ^k K^{i j} K_{j k}\\
 &+2\nabla ^cK^{a b} \left(\nabla _cK_{a b}-\nabla _bK_{a c}\right)+2 \omega ^c K^{a b} \nabla _cK_{a b}-2 \omega ^c K^{a b} \nabla _bK_{a c} -2\partial _vK K^{i j}\bar{K}_{i j}\\&+2\bar{K}K\partial_v K-2\bar{K}K^{i j}\partial _vK_{i j}-2K \bar{K}_{i j}\partial _vK^{i j}+4\partial_vK^{i j} \bar{K}_{j k}K_i^k \bigr]+\mathcal{O} (\epsilon ^3) \, .
\end{split}
\end{equation}
The $E_{vv}$ can be split into linear and non-linear pieces ($ E_{vv}(\epsilon) , E_{vv}(\epsilon^2)$). Hence, we may rewrite the definition of $F$ of eq.\eqref{eq:nonlinear} as
\begin{equation}\label{eq:f}
   F=  E_{vv}(\epsilon)+E_{vv}(\epsilon^2)-A(\epsilon) - B(\epsilon^2) \, .
 \end{equation}

\medskip

However, from eq.\eqref{eq:keyeqn} we know that for linearized approximation in $\epsilon$, 
\begin{equation}{\label{eq:g}}
    E_{vv}(\epsilon)= A (\epsilon) \, ,
\end{equation}
where the linear terms of $E_{vv}$ from eq.\eqref{eq:evv2} are given by
\begin{equation}\label{eq:evvlinear}
    E_{vv}(\epsilon)= 2 l^2 \left(2\mathcal{R}_{i j}\partial _vK^{i j}-\mathcal{R}\partial _vK\right) \, .
 \end{equation}
We have already proved in section \ref{sec:jvjitransfgen} that $A(\epsilon)$ transforms homogeneously up to linear order as per eq.\eqref{eq:jvjitransf2}. Hence, $E_{vv}(\epsilon)$ piece will also transform homogeneously up to linear order. However, one can check that the non-linear piece, i.e., $B(\epsilon^2)$ does not transform homogeneously:
\begin{equation}
	\begin{split}
		B-e^{-2 \zeta } \tilde{B}&= 4e^{-2 \zeta }\tilde{K}\left\{\left(\xi _i \xi _j \tilde{K}^{i j}-2\xi _i\tilde{\nabla }_j\tilde{K}^{i j}-\tilde{K}^{i j}\tilde{\nabla }_i\xi _j+2\xi ^i\tilde{\nabla }_i\tilde{K}-\xi ^2 \tilde{K}+\tilde{K}\tilde{\nabla }_i\xi ^i\right)\right.\\
		&+\tau \left(3\xi _i\xi _j\partial_{\tau }\tilde{K}^{i j}-2\xi _j\partial_{\tau }\tilde{\nabla }_i\tilde{K}^{i j}-\tilde{\nabla }_i\xi _j\partial _{\tau }\tilde{K}^{i j}+2\xi ^i\partial _{\tau }\tilde{\nabla }_i\tilde{K}-3\xi ^2\partial _{\tau }\tilde{K}+\tilde{\nabla }_i\xi ^i\partial _{\tau }\tilde{K} \right)\\
		&\left.+\tau ^2\left(\xi _i\xi _j\partial _{\tau }{}^2\tilde{K}^{i j}-\xi ^2\partial _{\tau }{}^2\tilde{K}\right)\right\}+\mathcal{O} (\epsilon ^3) \, .
	\end{split}
\end{equation}
This explicitly shows that for Einstein-Gauss-Bonnet theory, the `$E$' term of eq.\eqref{eq:lepbreak} does not transform homogeneously to non-linear order. From this, one can also conclude that $F$ of eq.\eqref{eq:nonlinear} (and hence eq.\eqref{eq:evvnonlinear}) does not transform homogeneously. The additional structure is that $E_{vv}(\epsilon^2)$ of eq.\eqref{eq:f} which consists of both $B$ and $F$ transforms homogeneously as eq.\eqref{eq:evvtrans}. Thus, there is a hierarchical structure to the $E_{vv}$ of eq.\eqref{eq:evvnonlinear}: $\mathcal{O}(\epsilon)$ terms transform homogeneously up to $\mathcal{O}(\epsilon^2)$ terms; $\mathcal{O}(\epsilon^2)$ terms transform homogeneously up to $\mathcal{O}(\epsilon^3)$ terms and so on.

\section{Discussions}
\label{sec:discussions}

Let us now briefly summarize what we have been able to achieve in our paper. We wanted to prove the covariance of entropy production explicated in eq.\eqref{eq:jvjitransf} under a reparametrization of the form eq.\eqref{eq:repara}. This result is explicitly checked in section \ref{sec:jvjitransfgen}. Consequently, we have derived eq.\eqref{eq:currentres} in eq.\eqref{eq:jvjifinaltransf}. eq.\eqref{eq:currentres} is an explicit formula that tells us how the components of the entropy current transform under a reparametrization of the form eq.\eqref{eq:repara}. In particular, the formula had non-homogeneous terms in addition to the homogeneous transformation pieces that represented the transformation of a covariant co-dimension one vector on the horizon. These non-homogeneous pieces can be interpreted as arising due to the intrinsic ambiguities of the form eq.\eqref{eq:ambinevv}. The effect of these ambiguities was detailed in section \ref{sec:ambcurrent}. They originated in the ambiguities arising from the covariant phase space quantities (namely $\Theta^{\mu}$  and $Q^{\mu\nu}$), and the final effect on the components of the entropy current are summarized in eq.\eqref{eq:finalambexpr}.  Following this in section \ref{sssec:ppfl}, we carefully analysed the physical process version of the first law for arbitrary time slices of the horizon. As a byproduct of this analysis, we naturally obtained an expression for the dynamical entropy of a black hole in a theory of the form eq.\eqref{eq:lag1}, which was derived to be eq.\eqref{eq:entropydyn}. Then, in section \ref{sec:verproof}, we explicitly verified the formulae eq.\eqref{eq:currentres}, thereby validating the abstract proof of section \ref{sec:jvjitransfgen}. Finally, in section \ref{sec:nonlinear}, we performed a preliminary analysis of the reparametrization covariance of entropy production to non-linear order in the dynamics. We observed that the entropy current structure $E$ of eq.\eqref{eq:evvnonlinear} doesn't transform homogeneously by itself. However, the non-covariant structures $E$ and $F$ of eq.\eqref{eq:evvnonlinear} together transform homogeneously as expected in eq.\eqref{eq:evvtrans}.

We now make some remarks on the possible future directions that could follow from our analysis. Although we were mainly motivated by extending the initial analysis of \cite{Bhattacharyya:2022njk} to arbitrary diffeomorphism invariant theories of gravity non-minimally coupled to matter fields, we also wanted to analyse the approach of \cite{Hollands:2022fkn} in proving the gauge invariance of the definition of a dynamical entropy for black holes in such theories. We emphasize that \cite{Hollands:2022fkn} was concerned with the question of the gauge invariance of the definition of entropy. In contrast, we are concerned with the gauge invariance of entropy production as in eq.\eqref{eq:jvjitransf}. The definition of entropy is given by the integral of $\mathcal{J}^v$ over a compact codimension two slice of the dynamical horizon. The approach of \cite{Hollands:2022fkn} argues that this definition is gauge invariant up to non-linear order in perturbations. However, in \cite{Davies:2022xdq}, it was shown that the gauge invariance will be lost if one considers Lagrangian theories with more than eight derivatives. Thus, it is useful to check whether our formalism of working with covariant structures like $E_{vv}$ to argue for the gauge invariance might help resolve this issue. Since the structure of eq.\eqref{eq:evvnonlinear} was argued in \cite{Hollands:2022fkn} using multiple components of EoM, a possible future direction is to consider different linear combinations of EoM and see if one can argue for gauge covariance (as EoM components transform homogeneously as in eq.\eqref{eq:evvtrans}).

One possible extension of this work would be to see if one can utilize the Iyer-Wald ambiguities to cancel out the non-homogeneous pieces in the transformation. It would be interesting to develop a systematic algorithm for such cancellations in arbitrary higher-derivative theories.

Another future direction is to extend our analysis to generic Chern-Simons theories of gravity considered in \cite{Deo:2023vvb}. The non-trivial point is that $\Theta^r$ and $Q^{r\mu}$ are no longer covariant and hence they do not transform like eq.\eqref{eq:thetartransf} and eq.\eqref{eq:qrvtransf} respectively under a reparametrization eq.\eqref{eq:repara}. Thus, we expect to have additional non-covariant terms in the final result of eq.\eqref{eq:jvjifinaltransf}.

The different coordinate systems of section \ref{ssec:difhorcoord} are related by affine transformations. It is interesting to study the consequences of exploring non-affine transformations that will take one away from the gauge of eq.\eqref{eq:metric1}. Certain aspects of these non-affine coordinate transformations have been studied in the context of fluid gravity correspondence in \cite{Chandranathan:2022pfx}. These transformations were the map relating the bulk horizon coordinates to the boundary fluid coordinates. It turns out that under this map, the boundary entropy current for Gauss-Bonnet theory is not solely dependent upon the fluid variables (like velocity, temperature and their derivatives) but also on the mapping functions between the affine coordinates on the horizon and the non-affine coordinates on the boundary. It would be interesting to see whether these ambiguities can be used to redefine the entropy currents in the affine coordinates so that they transform in a well-behaved manner to entropy currents of the boundary fluid, which depend only on fluid variables.

Finally, an interesting open direction would be the connection to the intrinsic BMS symmetries of the null horizon \cite{Donnay:2015abr,Blau:2015nee,Donnay:2016ejv,Grumiller:2019fmp,Adami:2021nnf,Adami:2023fbm}. In this language, eq.\eqref{eq:repara} should be thought of as a ``superrotation''. Now, there is an isomorphism between BMS symmetries in $d+1$ dimensions and conformal Carrollian symmetries in $d$ dimensions \cite{Duval:2014uva}. The transformation eq.\eqref{eq:currentres} is very reminiscent of how a Carrollian vector field transforms \cite{Ciambelli:2018ojf}. This is expected because, on the horizon, the metric degenerates. The precise connection between the non-covariant terms in eq.\eqref{eq:currentres} and the intrinsic Carrollian symmetries is worth exploring. It is also interesting to ask if this entropy current arises from some Carrollian description akin to \cite{Donnay:2019jiz}. We hope to address these questions in future works.

\section*{Acknowledgements}

We are greatly indebted to Sayantani Bhattacharyya and Nilay Kundu for valuable insights and collaboration on related ideas. We also thank Parthajit Biswas, Iain Davies, Krishna Jalan, Alok Laddha, Milan Patra, Harvey Reall, Amitabh Virmani, Aron Wall, Zihan Yan, and Mangesh Mandlik for useful discussions. 

\smallskip

We are grateful to the organizers and IIT Mandi for their hospitality at the 
``Student Talks in Trending Topics in Theory (ST4)'' workshop where some part of the work was carried out. We would like to thank the organizers of ISM 2023 for giving us an opportunity to present this work as a poster.  PD acknowledges the warm hospitality of University of Cambridge, University of Southampton, Durham University, King's College London, University of Edinburgh, IIT Bombay, Institute of Mathematical Sciences Chennai, Chennai Mathematical Institute, and BITS Goa during the course of this work. SR would like to acknowledge the warm hospitality of Chennai Mathematical Institute during the final stages of the work. PD would like to duly acknowledge the Council of Scientific and Industrial Research (CSIR), New Delhi, for financial assistance through the Senior Research Fellowship (SRF) scheme and the partial support from the Royal Society of London international exchange grant with the University of Edinburgh.

\appendix

\section{The basic setup and review of entropy current construction}
\label{ap:basicsetup}

In this Appendix, we will briefly review the essential elements that go into the construction of the entropy current \cite{Bhattacharyya:2021jhr}. For more details, the reader is referred to section 2 of \cite{Bhattacharyya:2021jhr}.

\subsubsection*{Residual gauge transformations in the gauge:}

The metric structure of the near horizon solution eq.\eqref{eq:metric1} is schematically described by figure \ref{fig:horizon}. This, however, will not fix the coordinates completely on constant $r$ and $v$ slices. We can do the following coordinate transformations while preserving the form of the gauge:
\begin{enumerate}
    \item Given two arbitrary functions $g_1(x^i)$ and $g_2(x^i)$, the transformation 
    \begin{equation}\label{eq:genvtransf}
        v \to \tilde{v} = g_1(x^i) \, v + g_2(x^i) \, ,
    \end{equation}
    with a suitable re-definition of $r$ leaves the metric eq.\eqref{eq:metric1} invariant. In section \ref{ssec:difhorcoord}, we will study this by explicitly constructing the functions $g_1(x^i)$ and $g_2(x^i)$. We will see that if one considers the form of the metric away from the horizon, these functions will, in principle, depend on the affine coordinates $r$ and $v$.
    \item A residual diffeomorphism invariance of the co-dimension two slice of the horizon
    \begin{equation}
        x^i \to \Tilde{x}^i = f^i(x^j) \, .
    \end{equation}
    This allows us to define a covariant derivative $\nabla_i$ compatible with the metric $h_{ij}$.
\end{enumerate}
The affine reparametrization eq.\eqref{eq:repara} is a special case of these residual coordinate transformations. 

\subsubsection*{Dynamical structures in the gauge:} As mentioned below eq.\eqref{eq:metric1}, we are interested in a dynamical situation near equilibrium. This is described by the perturbative approximation $g_{\mu\nu}=g^{eq}_{\mu\nu}+\epsilon \, \delta g_{\mu\nu}$, where $g^{eq}_{\mu\nu}$ is the equilibrium metric and $\epsilon \ll 1$ is the parameter characterizing the deviations from equilibrium. Thus, we need to construct an equilibrium solution with any available residual symmetry of the metric. We choose the following rescaling of coordinates to be the symmetry of the equilibrium metric:
\begin{equation}\label{eq:boosttransf}
    r \to \Tilde{r} = b \, r \, , ~~~ v \to \Tilde{v} = \dfrac{v}{b} \, ,
\end{equation}
where $b$ is a constant. This transformation is called `Boost transformation'. The equilibrium solution invariant under the boost transformation is given by
\begin{equation}\label{eq:eqmsol}
    ds^2 = 2 dv \, dr - r^2 \, X(rv,x^i) \, dv^2 + 2 r \, \omega_i(rv,x^i) \, dv \, dx^i + h_{ij}(rv,x^i) \, dx^i \, dx^j \, ,
\end{equation}
where the $v$ dependence of metric components is fixed through the product $rv$. The Killing vector that generates the horizon $r=0$ of the equilibrium solution is
\begin{equation}\label{eq:killing}
    \xi^{\alpha} \partial_{\alpha} = v \partial_v - r \partial_r \, .
\end{equation}
The perturbations about eq.\eqref{eq:eqmsol} are of the form
\begin{equation}\label{eq:pertubs}
    \begin{split}
        X(r,v,x^i) &= X^{eq}(rv,x^i) + \epsilon \, \delta X(r,v,x^i) \\
        \omega_i(r,v,x^i) &= \omega^{eq}_i(rv,x^i) + \epsilon \, \delta \omega_i(r,v,x^i) \\
        h_{ij}(r,v,x^i) &= h^{eq}_{ij}(rv,x^i) + \epsilon \,  \delta h_{ij}(r,v,x^i)
    \end{split}
\end{equation}

\medskip

Fixing the equilibrium solution allows us to classify the non-equilibrium structures using the notion of ``Boost weight''. To see this consider a quantity $\mathcal{B}$ that transforms in the following way under the boost transformation eq.\eqref{eq:boosttransf}:
\begin{equation}\label{eq:boostwt}
    \mathcal{B} \to \widetilde{\mathcal{B}} = b^w \mathcal{B} \, , ~~~ \text{under eq.\eqref{eq:boosttransf}} \, .
\end{equation}
$w$ is called the Boost weight of $\mathcal{B}$. If $w > 0$, then $\mathcal{B}$ is a dynamical structure on the horizon $r=0$. This is because $\mathcal{B}$ will be of the form $(\partial_r)^m (\partial_v)^{m+w} \mathcal{C}$ and $\mathcal{C}$ itself is given by
\begin{equation}
    \mathcal{C}(r,v,x^i) = \mathcal{C}^{eq}(rv,x^i) + \epsilon \, \delta \mathcal{C}(r,v,x^i) \, .
\end{equation}
This is because $\mathcal{C}$ is only constructed out of $X,\omega_i,h_{ij},\nabla_i$. Thus, we have
\begin{equation}
    (\partial_r)^m (\partial_v)^{m+w} \mathcal{C}|_{r=0} = \epsilon (\partial_r)^m (\partial_v)^{m+w} \delta \mathcal{C}|_{r=0} \sim \mathcal{O}(\epsilon) \, .
\end{equation}
From the boost weight criterion eq.\eqref{eq:boostwt}, it is clear that any covariant lower indexed tensor that has more number of $v$ indices when compared to $r$ indices is a dynamical quantity. The notion of boost weight in eq.\eqref{eq:boostwt} also gives additional structure to covariant quantities. For instance, according to ``Result: 1'' of \cite{Bhattacharyya:2021jhr}, a typical covariant tensor with boost weight $a+1$ on the horizon $r=0$ can be written as
\begin{equation}\label{eq:result1}
    \begin{split}
        t^{(k)}_{(a+1)}|_{r=0} &= M_{(-k)} \partial^{k+a+1}_v N_{(0)} + \mathcal{O}(\epsilon^2) \\
        &= \partial^{(a+1)}_v \left[ \sum^{k-1}_{m=0} (-1)^m \, [^{m+a}C_{m}] \, M_{(-k+m)}\partial^{k-m}_v N_{(0)} \right] \\
        & ~~ + (-1)^k \, [^{k+a}C_{a}] \,  M_{(0)} \partial^{(a+1)}_v N_{(0)} + \mathcal{O}(\epsilon^2) \, .
    \end{split}
\end{equation}

\subsubsection*{Local entropy production on the horizon:}

We will see how the structure of $E_{vv}$ described by eq.\eqref{eq:keyeqn} will lead to a local second law as eq.\eqref{eq:localseclaw}. Given the Lagrangian $\mathcal{L}$ of eq.\eqref{eq:lag1}, we add a null energy condition satisfying matter sector Lagrangian $\mathcal{L}_{\text{mat}}$. Thus, the full equation of motion (EoM) is given by
\begin{equation}\label{eq:eommunu}
    \mathcal{E}^{full}_{\mu\nu} = E_{\mu\nu} + T_{\mu\nu} \, ,
\end{equation}
where $E_{\mu\nu}$ is the contribution from eq.\eqref{eq:lag1} and $T_{\mu\nu}$ is the contribution from $\mathcal{L}_{\text{mat}}$. We now project the full EoM onto the null generators $\partial_v$ and then go on-shell. We, thus, have
\begin{equation}\label{eq:fullevvonshell}
    \mathcal{E}^{full}_{vv} = \partial_v \left( \dfrac{1}{\sqrt{h}} \partial_v \left( \sqrt{h} \, \mathcal{J}^v \right) + \dfrac{1}{\sqrt{h}}\partial_i \left( \sqrt{h} \, \mathcal{J}^i \right) \right) + T_{vv} + \mathcal{O}(\epsilon^2) = 0 \, ,
\end{equation}
where we have used eq.\eqref{eq:keyeqn}. The Null energy condition in our gauge eq.\eqref{eq:metric1} essentially amounts to $T_{vv} \geq 0$ and hence we have from eq.\eqref{eq:fullevvonshell}
\begin{equation}\label{eq:current1}
    \partial_v \left( \dfrac{1}{\sqrt{h}} \partial_v \left( \sqrt{h} \, \mathcal{J}^v \right) + \dfrac{1}{\sqrt{h}}\partial_i \left( \sqrt{h} \, \mathcal{J}^i \right) \right) \leq 0 ~~~ \text{upto} ~ \mathcal{O}(\epsilon^2) \, .
\end{equation}
We now use the physical boundary condition that the black hole settles down to some equilibrium solution in the far future as $v \to \infty$. Since $\mathcal{J}^v$ has boost weight $0$ and $\mathcal{J}^i$ has boost weight $+1$, we have
\begin{equation}\label{eq:current2}
    \dfrac{1}{\sqrt{h}} \partial_v \left( \sqrt{h} \, \mathcal{J}^v \right) + \dfrac{1}{\sqrt{h}}\partial_i \left( \sqrt{h} \, \mathcal{J}^i \right) \to 0 ~~~ \text{as} ~~~ v \to \infty \, .
\end{equation}
Using eq.\eqref{eq:current1} and eq.\eqref{eq:current2}, we straightforwardly have
\begin{equation}\label{eq:aplocseclaw}
    \dfrac{1}{\sqrt{h}} \partial_v \left( \sqrt{h} \, \mathcal{J}^v \right) + \dfrac{1}{\sqrt{h}}\partial_i \left( \sqrt{h} \, \mathcal{J}^i \right) \geq 0 ~ \text{upto} ~ \mathcal{O}(\epsilon^2) ~~~ \text{for all finite $v$} \, .
\end{equation}
This signifies ultra-local entropy production on the horizon as in eq.\eqref{eq:localseclaw}.

\subsubsection*{Proof via an off-shell identity:}

We will now briefly sketch the proof of the structure of $E_{vv}$ (eq.\eqref{eq:keyeqn}) for arbitrary higher derivative theories of gravity. Certain elements of the proof will be crucial for the analysis in sections \ref{sec:jvjitransfgen} and \ref{sec:ambcurrent}. We will use the structures of the covariant phase space formalism \cite{} to facilitate this. For the sake of simplicity, we will only review the case of pure gravitational Lagrangian of the form
\begin{equation}\label{eq:lpure}
    L = L(g_{\mu\nu},R_{\mu\nu\rho\sigma},D_{\alpha_1}R_{\mu\nu\rho\sigma},\dots) \, .
\end{equation}
The variation of the Lagrangian $L$ of eq.\eqref{eq:lpure} with respect to the metric is given by
\begin{equation}\label{eq:varL}
    \delta [\sqrt{-g} L]=\sqrt{-g} E^{\mu\nu}_g \delta g_{\mu\nu} + \sqrt{-g} D_\mu \Theta^\mu [\delta g] \, .
\end{equation}
The crucial off-shell identity comes about if we consider a variation that is produced due to an infinitesimal diffeomorphism $x^{\mu} \to x^{\mu} + \zeta^{\mu}$ (i.e.) $\delta g_{\mu\nu} = D_{\mu} \zeta_{\nu} + D_{\nu} \zeta_{\mu}$ for which we have
\begin{equation}\label{eq:Ldiff}
    \delta \left[ \sqrt{-g} L \right] = \sqrt{-g} D_{\mu} (\zeta^{\mu} L) \, . 
\end{equation}
Substituting eq.\eqref{eq:Ldiff} in eq.\eqref{eq:varL} and using the Generalized Bianchi identity $D_{\mu} E^{\mu\nu} = 0$ \cite{10.1063/1.528801}, we have
\begin{equation}
    D_{\mu} \left( \zeta^{\mu} L - 2 E^{\mu\nu} \zeta_{\nu} - \Theta^{\mu}[\mathcal{L}_{\xi} g_{\mu\nu}] \right) = 0 \, .
\end{equation}
Since we have a conserved vector, we can express the above equation in terms of the ``Iyer-Wald'' charge $Q^{\mu\nu}$
\begin{equation}\label{eq:jmu}
    \zeta^{\mu} L - 2 E^{\mu\nu} \zeta_{\nu} - \Theta^{\mu}[\mathcal{L}_{\xi} g_{\mu\nu}] = - D_{\nu} Q^{\mu\nu} \, ,
\end{equation}
for an antisymmetric $Q^{\mu\nu} = - Q^{\nu\mu}$. The equation eq.\eqref{eq:jmu} is true for any $\zeta^{\mu}$ but we will choose $\zeta^{\mu} = \xi^{\mu}$ given by the Killing vector of the equilibrium solution eq.\eqref{eq:killing}. Contracting eq.\eqref{eq:jmu} with $\xi^{\mu}$ on $r=0$, we finally get
\begin{equation}\label{eq:mainrel}
    2 v \, E_{vv}|_{r=0} = \left( - \Theta^r + D_{\mu} Q^{r\mu}\right)|_{r=0} \, .
\end{equation}
This is an important equation in the analysis of \cite{Bhattacharyya:2021jhr,Biswas:2022grc} because while we have no general structures for the equations of motion for arbitrary higher derivative theories of gravity, we have general structures for $\Theta^r$ and $Q^{r\mu}$ \cite{Iyer:1994ys}. Using the general structures, one can show that in our gauge eq.\eqref{eq:metric1} they take the following form on the horizon $r=0$:
\begin{equation}\label{eq:thetaqfinal}
     \begin{split}
         \Theta^r|_{r=0} &= (1+v\partial_v) \mathcal{A}_{(1)} + v\partial^2_v \mathcal{B}_{(0)} + \mathcal{O}(\epsilon^2) \, , \\
         Q^{r\mu} &= \mathcal{Q}^{r\mu}+v \, W^{r\mu}_v \, .
     \end{split}
\end{equation}
Substituting eq.\eqref{eq:thetaqfinal} in eq.\eqref{eq:mainrel}, we have
\begin{equation}\label{eq:preevv}
    \begin{split}
        2v \, E_{vv} = &\dfrac{1}{\sqrt{h}} \partial_v \left( \sqrt{h} \, \mathcal{Q}^{rv} \right) + \nabla_i \mathcal{Q}^{ri} + W^{rv}_v - \mathcal{A}_{(1)} \\
        & +v \partial_v \left[ W^{rv}_v + \nabla_i J^i_{(1)} -\mathcal{A}_{(1)} - \partial_v \mathcal{B}_{(0)} \right] + \mathcal{O}(\epsilon^2) \, .
    \end{split}
\end{equation}
In writing the second line of eq.\eqref{eq:preevv}, we have used
\begin{equation}
    W^{ri}_v = \partial_v J^i_{(1)} + \mathcal{O}(\epsilon^2) \implies \nabla_i W^{ri}_v = \partial_v \left( \nabla_i J^i_{(1)} \right) + \mathcal{O}(\epsilon^2) \, ,
\end{equation}
which is a consequence of eq.\eqref{eq:result1}. Since then we have explicated the $v$ dependence in eq.\eqref{eq:preevv}, we can explicitly compare the coefficients of $v^0$ and $v^1$ in eq.\eqref{eq:preevv} to write
\begin{equation}\label{eq:preevv2}
    \begin{split}
        \mathcal{A}_{(1)} &= \dfrac{1}{\sqrt{h}} \partial_v \left( \sqrt{h} \, \mathcal{Q}^{rv} \right) + \nabla_i \mathcal{Q}^{ri} + W^{rv}_v \, , \\
        2E_{vv} &= \partial_v \left[W^{rv}_v + \nabla_i J^i_{(1)} -\mathcal{A}_{(1)} - \partial_v \mathcal{B}_{(0)} \right] \, .
    \end{split}
\end{equation}
eq.\eqref{eq:preevv} is justified in Appendix A.1.1 of \cite{Deo:2023vvb} and this has also been emphasized in \cite{Hollands:2022fkn}. Substituting for $\mathcal{A}_{(1)}$ from the first equation of eq.\eqref{eq:preevv2} into the second equation, we have
\begin{equation}\label{eq:evvapfinal}
    2 E_{vv} |_{r=0} = -\partial_v \left[\dfrac{1}{\sqrt{h}} \partial_v \left( \sqrt{h} \left(\mathcal{Q}^{rv} + \mathcal{B}_{(0)}\right) \right) + \nabla_i \left[ \mathcal{Q}^{ri} - J^i_{(1)} \right]\right] + \mathcal{O}(\epsilon^2) \, ,
\end{equation}
where $\nabla_i$ is the covariant derivative associated with $h_{ij}$. Thus, we have proved eq.\eqref{eq:keyeqn}, concluding our brief review of the construction of entropy current. The generalization to $U(1)$ gauge theories requires more technical machinery described in \cite{Biswas:2022grc} to bring the structure of $\Theta^r$ and $Q^{r\mu}$ to the form of eq.\eqref{eq:thetaqfinal}. We will be using eq.\eqref{eq:thetaqfinal} and eq.\eqref{eq:evvapfinal} to argue for the gauge covariance of local entropy production in section \ref{sec:jvjitransfgen}.

\section{Intricate details of the computation}
\label{ap:detailscomp}
In this Appendix, we collect the details of the computations made in section \ref{sec:jvjitransfgen}. For the computations, we keep terms only up to $\mathcal{O}(\epsilon) \,$.

\begin{itemize}
  \item Transformation of $\mathcal{Q}^{ri}$ and 
       $ W^{ri}_v$ 
      \begin{equation}\label{eq:apqritransf} 
       \begin{split}
           \mathcal{Q}^{ri} &= 4 E^{rirv}_R \\&=4 \frac{\partial r}{\partial \rho}\frac{\partial x^{i}}{\partial y^{a}}\frac{\partial r}{\partial \rho}\frac{\partial v}{\partial \tau}  \widetilde{E}^{\rho a \rho \tau}_R +4\frac{\partial r}{\partial \rho}\frac{\partial x^{i}}{\partial y^{a}}\frac{\partial r}{\partial \rho}\frac{\partial v}{\partial y^b}  \widetilde{E}^{\rho a \rho b}_R\\&=4 e^{-\zeta} \widetilde{E}^{\rho i \rho \tau}_R + 4 e^{-\zeta} \tau \xi_b \widetilde{E}^{\rho i \rho b}_R \\&= e^{-\zeta} \widetilde{\mathcal{Q}}^{\rho i } + 4 e^{-\zeta} \tau \xi_b \widetilde{E}^{\rho i \rho b}_R  
       \end{split}
      \end{equation}

      \begin{equation}\label{eq:apwrivtransf} 
       \begin{split}
           W^{ri}_v &= -2 E^{rirj}_R \omega_j+W^{rir} \\&=-2 \frac{\partial r}{\partial \rho}\frac{\partial x^{i}}{\partial y^{a}}\frac{\partial r}{\partial \rho}\frac{\partial x^j}{\partial y^b}  \widetilde{E}^{\rho a \rho b}_R \omega_b +\frac{\partial r}{\partial \rho}\frac{\partial x^{i}}{\partial y^{a}}\frac{\partial r}{\partial \rho} \widetilde{W}^{\rho a \rho }\\&= -2e^{-2\zeta} \widetilde{E}^{\rho i \rho b}_R (\widetilde{\omega}_b+ 2 \xi _b - 2 \tau  \xi _k \widetilde{h}^{jk} \widetilde{K}_{bj}) +e^{-2\zeta}\widetilde{W}^{\rho i \rho }\\&=-2e^{-2\zeta}\widetilde{E}^{\rho i \rho b}_R \widetilde{\omega}_b-4 e^{-2\zeta} \xi_b \widetilde{E}^{\rho i \rho b}_R+e^{-2\zeta} \widetilde{W}^{\rho i \rho}\\
           &= e^{-2\zeta} \widetilde{W}^{\rho i}_{\tau} - 4 e^{-2\zeta} \xi_b \widetilde{E}^{\rho i \rho b}_R .  
       \end{split}
      \end{equation}
      These equations prove eq.\eqref{eq:qriw}.

      \item Transformation of $J^i_{(1)}$

      By definition of eq.\eqref{eq:qwexpr} $W^{ri}_v=\partial_v J^i_{(1)} + \mathcal{O}(\epsilon^2)$. As obtained above,
       \begin{equation}\label{eq:apji1transf}
           \begin{split}
             &W^{ri}_v =  e^{-2\zeta} \widetilde{W}^{\rho i}_{\tau} - 4 e^{-2\zeta} \xi_b \widetilde{E}^{\rho i \rho b}_R \\
             &\implies \partial_v {J^i_{(1)}}= e^{-2\zeta} \partial_\tau {\widetilde{J}^i_{(1)}}- 4 e^{-2\zeta} \xi_b \widetilde{E}^{\rho i \rho b}_R\\
             &\implies e^{-\zeta}\partial_\tau {J^i_{(1)}}= e^{-2\zeta} \partial_\tau {\widetilde{J}^i_{(1)}}- 4 e^{-2\zeta} \xi_b \partial_\tau{\widetilde{L}^{ i b}_{(1)}}\\
             &\implies {J^i_{(1)}}= e^{-\zeta}{\widetilde{J}^i_{(1)}}- 4 e^{-\zeta} \xi_b \widetilde{L}^{ i b}_{(1)}
           \end{split}
       \end{equation}
       This proves eq.\eqref{eq:ji1transf}.

        \item Transformation of $\mathcal{J}^i$
       \begin{equation}\label{eq:apjitransf}
           \begin{split}
            \mathcal{J}^i &= -\dfrac{1}{2} \left( \mathcal{Q}^{ri} -  J^i_{(1)} \right) \, .\\
            &=-\dfrac{1}{2}(e^{-\zeta} \widetilde{\mathcal{Q}}^{\rho i } + 4 e^{-\zeta} \tau \xi_b \widetilde{E}^{\rho i \rho b}_R-e^{-\zeta}{\widetilde{J}^i_{(1)}}+ 4 e^{-\zeta} \xi_b \widetilde{L}^{ i b}_{(1)})\\&=-\dfrac{1}{2}(e^{-\zeta} \widetilde{\mathcal{Q}}^{\rho i } + 4 e^{-\zeta} \tau \xi_b \partial_\tau{\widetilde{L}^{ i b}_{(1)}} -e^{-\zeta}{\widetilde{J}^i_{(1)}}+ 4 e^{-\zeta} \xi_b \widetilde{L}^{ i b}_{(1)})\\&=-\dfrac{1}{2}(e^{-\zeta} \widetilde{\mathcal{Q}}^{\rho i } -e^{-\zeta}{\widetilde{J}^i_{(1)}})-2 e^{-\zeta} (1+\tau\partial_\tau) (\xi_b\widetilde{L}^{ i b}_{(1)})\\
            &=e^{-\zeta} \widetilde{\mathcal{J}}^{i } -2 e^{-\zeta} (1+\tau\partial_\tau) (\xi_j\widetilde{L}^{ i j}_{(1)})\\
           \end{split}
       \end{equation}
       This proves eq.\eqref{eq:jifinaltransf}.
 \item Transformation of $\mathcal{Q}^{rv}$ and 
       $ W^{rv}_v$ 
      \begin{equation}\label{eq:apvarqrvtransf} 
       \begin{split}
           \mathcal{Q}^{rv} &= 4 E^{rvrv}_R \\&=4 \frac{\partial r}{\partial \rho}\frac{\partial v}{\partial \tau} \frac{\partial r}{\partial \rho}\frac{\partial v}{\partial \tau}  \widetilde{E}^{\rho \tau \rho \tau}_R +4\frac{\partial r}{\partial \rho}\frac{\partial v }{\partial y^{i}}\frac{\partial r}{\partial \rho}\frac{\partial v}{\partial \tau}  \widetilde{E}^{\rho i \rho \tau}_R+4\frac{\partial r}{\partial \rho}\frac{\partial v} {\partial \tau}\frac{\partial r}{\partial \rho}\frac{\partial v}{\partial y^i} \widetilde{E}^{\rho \tau \rho i}_R\\&+4\frac{\partial r}{\partial \rho}\frac{\partial v} {\partial y^{i}}\frac{\partial r}{\partial \rho}\frac{\partial v}{\partial y^j}  \widetilde{E}^{\rho i \rho j}_R \\&=4 \widetilde{E}^{\rho \tau \rho \tau}_R+8\tau \xi_i \widetilde{E}^{\rho i \rho \tau}_R+4 \tau^2 \xi_i \xi_j \widetilde{E}^{\rho i \rho j }_R\\&=\widetilde{\mathcal{Q}}^{\rho\tau}+2\xi_i\tau \widetilde{\mathcal{Q}}^{\rho i}+4 \tau^2 \xi_i \xi_j \widetilde{E}^{\rho i \rho j} 
       \end{split}
      \end{equation}
      This proves eq.\eqref{eq:qtilderv}.

 \begin{equation}\label{eq:apwrvvtransf} 
       \begin{split}
           W^{rv}_v &= -2 E^{rirv}_R \omega_i+W^{rvr} \\&=\frac{\partial r}{\partial \rho}\frac{\partial v}{\partial \tau}\frac{\partial r}{\partial \rho}\widetilde{W}^{\rho\tau\rho}+ \frac{\partial r}{\partial \rho}\frac{\partial v}{\partial y^i}\frac{\partial r}{\partial \rho}\widetilde{W}^{\rho i \rho}-2\frac{\partial r}{\partial \rho}\frac{\partial x^i}{\partial y^a}\frac{\partial r}{\partial \rho}\frac{\partial v}{\partial \tau}  \widetilde{E}^{\rho a \rho \tau}_R (\widetilde{\omega_i}+2\xi_i)\\& -2\frac{\partial r}{\partial \rho}\frac{\partial x^{i}}{\partial y^{a}}\frac{\partial r}{\partial \rho}\frac{\partial v}{\partial y^b}\widetilde{E}^{\rho a \rho b}_R (\widetilde{\omega_i}+2\xi_i)+\mathcal{O}(\epsilon^2) \, \\&=e^{-\zeta}\widetilde{W}^{\rho\tau\rho}+\tau e^{-\zeta}\xi_i\widetilde{W}^{\rho i \rho} -2 e^{-\zeta} \widetilde{E}^{\rho i \rho \tau}_R (\widetilde{\omega_i}+2\xi_i)-2 e^{-\zeta}\tau \xi_b \widetilde{E}^{\rho i \rho b}_R (\widetilde{\omega_i}+2\xi_i)\\&=e^{-\zeta}(\widetilde{W}^{\rho\tau\rho}-2\widetilde{E}^{\rho i \rho \tau}_R \widetilde{\omega_i})+\tau e^{-\zeta}\xi_i(\widetilde{W}^{\rho i \rho} -2 \widetilde{E}^{\rho i \rho \tau}_R \widetilde{\omega_b})-4 e^{-\zeta} \xi_i \widetilde{E}^{\rho i \rho \tau}_R\\&-4 \tau e^{-\zeta}\xi_i\xi_j\widetilde{E}^{\rho i \rho j}_R\\&=e^{-\zeta}(\widetilde{W}^{\rho\tau\rho}-2\widetilde{E}^{\rho i \rho \tau}_R \widetilde{\omega_i})+\tau e^{-\zeta}\xi_i(\widetilde{W}^{\rho i \rho} -2 \widetilde{E}^{\rho i \rho b}_R \widetilde{\omega_b})-4 e^{-\zeta} \xi_i \widetilde{E}^{\rho i \rho \tau}_R\\&-4 \tau e^{-\zeta}\xi_i\xi_j \widetilde{E}^{\rho i \rho j}_R +\\&= e^{-\zeta}\left(\widetilde{W}^{\rho \tau}_\tau +\tau \xi_i \widetilde{W}^{\rho i}_\tau -4\xi_i \widetilde{E}^{\rho i \rho \tau}_R-4\tau\xi_i\xi_j\widetilde{E}^{\rho i \rho j}_R\right),
       \end{split}
      \end{equation}
      This proves eq.\eqref{eq:wrvv}.

 \item Transformation of $Q^{rv}$
 \begin{equation}\label{eq:apqrvtransf}
     \begin{split}
       Q^{rv} &= \mathcal{Q}^{rv}+vW^{rv}_v\\&=\widetilde{\mathcal{Q}}^{\rho\tau}+2\xi_i\tau \widetilde{\mathcal{Q}}^{\rho i}+4 \tau^2 \xi_i \xi_j \widetilde{E}^{\rho i \rho j} \\ & ~+ \tau e^\zeta \left[e^{-\zeta}\left(\widetilde{W}^{\rho \tau}_\tau +\tau \xi_i \widetilde{W}^{\rho i}_\tau -4\xi_i \widetilde{E}^{\rho i \rho \tau}_R-4\tau\xi_i\xi_j\widetilde{E}^{\rho i \rho j}_R\right)\right]\\
       &=\widetilde{\mathcal{Q}}^{\rho\tau}+\tau\widetilde{W}^{\rho \tau}_\tau+ 2\xi_i\tau \widetilde{\mathcal{Q}}^{\rho i}-\tau4\xi_i \widetilde{E}^{\rho i \rho \tau}_R\\&=\widetilde{Q}^{\rho\tau}+2\xi_i\tau \widetilde{\mathcal{Q}}^{\rho i}-\tau \xi_i \widetilde{\mathcal{Q}}^{\rho i}\\&=\widetilde{Q}^{\rho\tau}+\tau \xi_i \widetilde{Q}^{\rho i} 
     \end{split}
 \end{equation}
 This proves eq.\eqref{eq:qrvtransf}.

 \item Transformation of $\mathcal{A}_{(1)}$. \\
 By the definition of eq.\eqref{eq:preevv2}
\begin{equation}\label{eq:apa1transf}
    \begin{split}
        \mathcal{A}_{(1)} &= \dfrac{1}{\sqrt{h}}\partial_v \left( \sqrt{h} \, \mathcal{Q}^{rv} \right) + \dfrac{1}{\sqrt{h}} \partial_i \left( \sqrt{h} \, \mathcal{Q}^{ri} \right) + W^{rv}_v \, \\&=
        \dfrac{1}{\sqrt{\widetilde{h}}}\partial_v \left( \sqrt{\widetilde{h}} \,(\widetilde{\mathcal{Q}}^{\rho\tau}+2\xi_i\tau \widetilde{\mathcal{Q}}^{\rho i}+4 \tau^2 \xi_i \xi_j \widetilde{E}^{\rho i \rho j}_R)\right)+\dfrac{1}{\sqrt{\widetilde{h}}}(\widetilde{\partial_i}-\tau \xi_i\partial_{\tau})\\&\times \left(\sqrt{\widetilde{h}}(e^{-\zeta} \widetilde{\mathcal{Q}}^{\rho i }+ 4 e^{-\zeta} \tau \xi_b \widetilde{E}^{\rho i \rho b}_R)\right)+e^{-\zeta}\left(\widetilde{W}^{\rho \tau}_\tau +\tau \xi_i \widetilde{W}^{\rho i}_\tau -4\xi_i \widetilde{E}^{\rho i \rho \tau}_R-4\tau\xi_i\xi_j\widetilde{E}^{\rho i \rho j}_R\right)\\&=\dfrac{1}{\sqrt{\widetilde{h}}}e^{-\zeta}\partial_{\tau}(\sqrt{\widetilde{h}}\widetilde {\mathcal{Q}}^{\rho \tau} )+2 e^{-\zeta} \xi_i \widetilde{\mathcal{Q}}^{\rho i} + 2e^{-\zeta}\tau \xi_i\partial_\tau \widetilde {\mathcal{Q}}^{\rho i} +8 e^{-\zeta} \tau \xi_i\xi_j \widetilde{E}^{\rho i \rho j}_R\\&+4 \tau^2 e^{-\zeta} \xi_i\xi_j \partial_\tau \widetilde{E}^{\rho i \rho j}_R+ \dfrac{1}{\sqrt{\widetilde{h}}}e^{-\zeta}\widetilde{\partial}_{i}(\sqrt{\widetilde{h}}\widetilde {\mathcal{Q}}^{\rho i} )-e^{-\zeta} \xi_i \widetilde{\mathcal{Q}}^{\rho i}-e^{-\zeta}\tau \xi_i\partial_\tau \widetilde {\mathcal{Q}}^{\rho i}\\&+\dfrac{1}{\sqrt{\widetilde{h}}}e^{-\zeta}\widetilde{\partial}_i (\sqrt{\widetilde{h}}4\tau \xi_b{\partial}_{\tau}\widetilde {L}^{i b}_{(1)} )-4\tau  e^{-\zeta} \xi_i\xi_b \widetilde{E}^{\rho i \rho b}_R+ e^{-\zeta}\widetilde{W}^{\rho \tau}_\tau+e^{-\zeta}\tau\xi_i\widetilde{W}^{\rho i}_\tau \\&- e^{-\zeta}\xi_i\widetilde {\mathcal{Q}}^{\rho i}-4\tau  e^{-\zeta} \xi_i\xi_j \partial_\tau \widetilde{E}^{\rho i \rho j}_R-4\tau  e^{-\zeta} \xi_i\xi_j \partial_\tau \widetilde{E}^{\rho i \rho j}_R-4 \tau^2 e^{-\zeta} \xi_i\xi_j \partial_\tau \widetilde{E}^{\rho i \rho j}_R\\&=
        e^{-\zeta}\left(\dfrac{1}{\sqrt{\widetilde{h}}}e^{-\zeta}\partial_{\tau}+\dfrac{1}{\sqrt{\widetilde{h}}}e^{-\zeta}\widetilde{\partial}_{i}(\sqrt{\widetilde{h}}\widetilde {\mathcal{Q}}^{\rho i} )+\widetilde{W}^{\rho \tau}_\tau\right)+e^{-\zeta}\tau \xi_i\partial_\tau \widetilde {\mathcal{Q}}^{\rho i}\\&+e^{-\zeta}\tau \widetilde{\nabla}_i(4\xi_j \partial_{\tau}\widetilde{L}^{ij}_{(1)})+e^{-\zeta}\tau \xi_i\partial_{\tau} \widetilde{J}^i_{(1)}-4e^{-\zeta}\tau \xi_i \xi_j \partial_{\tau} \widetilde{L}^{ij}_{(1)}  \\&=e^{-\zeta} \widetilde{\mathcal{A}}_{(1)} + e^{-\zeta} \tau \xi_i \partial_{\tau} \widetilde{\mathcal{Q}}^{\rho i} + e^{-\zeta} \tau \widetilde{\nabla_i} \left( 4 \xi_j \partial_{\tau} \widetilde{L}^{ij}_{(1)} \right) + e^{-\zeta} \tau \xi_i \partial_{\tau} \widetilde{J}^i_{(1)} \\&- 4 e^{-\zeta} \tau \xi_i \xi_j \partial_{\tau} \widetilde{L}^{ij}_{(1)} .
    \end{split}
\end{equation}
This proves eq.\eqref{eq:a1finaltransf}.

 \item Transformation of $\mathcal{B}_{(0)}$ 

 We begin with the expression of $\Theta^r$ on the horizon $r=0$ i.e,
\begin{equation}
   \begin{split}
    \Theta^r|_{r=0}&=(1+v\partial_v) \mathcal{A}_{(1)} + v\partial^2_v \mathcal{B}_{(0)}\\
    &=e^{-\zeta}(1+\tau\partial_{\tau}) \widetilde{\mathcal{A}}_{(1)} + e^{-\zeta}\tau\partial^2_\tau {\mathcal{B}}_{(0)}+e^{-\zeta}(1+\tau\partial_{\tau})(\tau\xi_i\partial_{\tau}\widetilde{\mathcal{Q}}^{\rho i})\\&+e^{-\zeta}(1+\tau\partial_{\tau})(\tau\xi_i\partial_{\tau}\widetilde{{J}}^i_{(1)})+e^{-\zeta}(1+\tau\partial_{\tau})\left(\tau\widetilde{\nabla_i} \left( 4 \xi_j \partial_{\tau} \widetilde{L}^{ij}_{(1)} \right)\right)\\&-e^{-\zeta}(1+\tau\partial_{\tau})\left(-4\tau\xi_i\xi_j\partial_{\tau}\widetilde{L}^{i j}_{(1)}\right)\\&=e^{-\zeta}(1+\tau\partial_{\tau}) \widetilde{\mathcal{A}}_{(1)} +e^{-\zeta}\partial^2_\tau {\mathcal{B}}_{(0)}+e^{-\zeta}\tau(2+\tau\partial_{\tau})\left(\xi_i\partial_{\tau}\widetilde{\mathcal{Q}}^{\rho i}+\xi_i\partial_{\tau}\widetilde{{J}}^i_{(1)} \right. \\
    & \hspace{3cm} \left. +\widetilde{\nabla_i} ( 4 \xi_j \partial_{\tau} \widetilde{L}^{ij}_{(1)})-4\xi_i\xi_j\partial_{\tau}\widetilde{L}^{i j}_{(1)}\right)\\    
    &\equiv e^{-\zeta}(1+\tau\partial_{\tau}) \widetilde{\mathcal{A}}_{(1)} + e^{-\zeta}\tau\partial^2_\tau \widetilde{ {\mathcal{B}}}_{(0)}
    \end{split}
\end{equation}

 where  \\
 $ \partial^2_\tau  \widetilde{ {\mathcal{B}}}_{(0)}=\partial^2_\tau {\mathcal{B}}_{(0)}+(2+\tau\partial_{\tau})\left(\xi_i\partial_{\tau}\widetilde{\mathcal{Q}}^{\rho i}+\xi_i\partial_{\tau}\widetilde{{J}}^i_{(1)}+\widetilde{\nabla_i} ( 4 \xi_j \partial_{\tau} \widetilde{L}^{ij}_{(1)})-4\xi_i\xi_j\partial_{\tau}\widetilde{L}^{i j}_{(1)}\right)$

\begin{equation*}
    \begin{split}
     \partial^2_\tau  \widetilde{ {\mathcal{B}}}_{(0)}&=\partial^2_\tau {\mathcal{B}}_{(0)}+(2+\tau\partial_{\tau})\left(\xi_i\partial_{\tau}\widetilde{\mathcal{Q}}^{\rho i}+\xi_i\partial_{\tau}\widetilde{{J}}^i_{(1)}+\widetilde{\nabla_i} ( 4 \xi_j \partial_{\tau} \widetilde{L}^{ij}_{(1)})-4\xi_i\xi_j\partial_{\tau}\widetilde{L}^{i j}_{(1)}\right)   \\ \partial^2_v   {\mathcal{B}}_{(0)}&=e^{-2\zeta}\left(\partial^2_\tau \widetilde{{\mathcal{B}}}_{(0)}-(2+\tau\partial_{\tau})\left(\xi_i\partial_{\tau}\widetilde{\mathcal{Q}}^{\rho i}+\xi_i\partial_{\tau}\widetilde{{J}}^i_{(1)}+\widetilde{\nabla_i} ( 4 \xi_j \partial_{\tau} \widetilde{L}^{ij}_{(1)})-4\xi_i\xi_j\partial_{\tau}\widetilde{L}^{i j}_{(1)}\right) \right)  
    \end{split}
\end{equation*}

 Considering,

$(2+\tau\partial_{\tau})(\xi_i\partial_{\tau}\widetilde{\mathcal{Q}}^{\rho i})=\partial_{\tau}\left((1+\tau\partial_{\tau})(\xi_i \widetilde{\mathcal{Q}}^{\rho i})\right)=\partial_{\tau}\left(\partial_{\tau}(\tau \xi_i \widetilde{\mathcal{Q}}^{\rho i})\right)$

We have,
\begin{equation}\label{eq:apbotransf}
    \begin{split}
     \partial^2_v {\mathcal{B}}_{(0)}&= e^{-2\zeta}\left(\partial^2_\tau \widetilde{{\mathcal{B}}}_{(0)}-\partial^2_{\tau}\left(\tau\xi_i\widetilde{\mathcal{Q}}^{\rho i}+\tau\xi_i\widetilde{{J}}^i_{(1)}+\tau\widetilde{\nabla_i} ( 4 \xi_j \widetilde{L}^{ij}_{(1)})-4\tau \xi_i\xi_j\widetilde{L}^{i j}_{(1)}\right) \right)  \\&=e^{-2\zeta}\partial^2_\tau \widetilde{\mathcal{B}}_{(0)}
\\\implies {\mathcal{B}}_{(0)}&=\widetilde{\mathcal{B}}_{(0)}-   \left(\tau \xi_i\widetilde{\mathcal{Q}}^{\rho i}+\tau\xi_i\widetilde{{J}}^i_{(1)}+\tau\widetilde{\nabla_i} ( 4 \xi_j \widetilde{L}^{ij}_{(1)})-4\tau\xi_i\xi_j\widetilde{L}^{i j}_{(1)}\right)   
    \end{split}
\end{equation}
This proves eq.\eqref{eq:botransf}.

\item Transformation of $\mathcal{J}^v$ 
\begin{equation}\label{eq:apjvtransf}
    \begin{split}
     &\mathcal{J}^v = -\dfrac{1}{2}\left( \mathcal{Q}^{rv} + \mathcal{B}_{(0)} \right)\\&= -\dfrac{1}{2}\bigg [\widetilde{\mathcal{Q}}^{\rho\tau}+2\xi_i\tau \widetilde{\mathcal{Q}}^{\rho i}+4 \tau^2 \xi_i \xi_j \widetilde{E}^{\rho i \rho j} + \widetilde{\mathcal{B}}_{(0)}-   \bigg(\tau \xi_i\widetilde{\mathcal{Q}}^{\rho i}+\tau\xi_i\widetilde{{J}}^i_{(1)}+\tau\widetilde{\nabla_i} ( 4 \xi_j \widetilde{L}^{ij}_{(1)})\\&-4\tau\xi_i\xi_j\widetilde{L}^{i j}_{(1)}\bigg) \bigg] \\&= \widetilde{\mathcal{J}}^{\tau}-\dfrac{1}{2}\left(\xi_i\tau \widetilde{\mathcal{Q}}^{\rho i}-\tau\xi_i\widetilde{{J}}^i_{(1)}+4 \tau^2 \xi_i \xi_j \partial_{\tau}\widetilde{L}^{i j}_{(1)} -\tau\widetilde{\nabla_i} ( 4 \xi_j \widetilde{L}^{ij}_{(1)})+4\tau\xi_i\xi_j\widetilde{L}^{i j}_{(1)}\right)\\&=\widetilde{\mathcal{J}}^{\tau} + \tau \xi_i \widetilde{\mathcal{J}}^i - 2 \tau^2 \xi_i \xi_j \partial_{\tau} \widetilde{L}^{ij}_{(1)} + 2 \tau \widetilde{\nabla}_i \left( \xi_j \widetilde{L}^{ij}_{(1)} \right) - 2 \tau \xi_i \xi_j \widetilde{L}^{ij}_{(1)} \, .
    \end{split}
\end{equation}
This proves eq.\eqref{eq:jvtransf}.

\end{itemize}

\section{Change of structures under reparametrization}
\label{ap:reparadetails}

In this Appendix, we will list the transformation of various structures under a reparametrization. On the horizon, the Christoffel symbols corresponding to the induced metrics $h_{ij}$ and $\widetilde{h}_{ij}$ are related by
\begin{equation}
    \Gamma _{a,b c}(h)=\widetilde{\Gamma }_{a,b c}(\widetilde{h})-\tau  \left(\xi _c \widetilde{K}_{a b}+\xi _b \widetilde{K}_{a c}-\xi _a \widetilde{K}_{b c}\right) \, ,
\end{equation}

where the extrinsic curvatures $(K, \bar{K})$ and $(\widetilde{K}, \widetilde{\bar{K}})$ are defined as
\begin{align}
    K_{ij} &= \frac{1}{2} \partial_v h_{ij} & K &= h^{ij} K_{ij} & \widetilde{K}_{ij} &= \frac{1}{2} \partial_\tau \widetilde{h}_{ij} & \widetilde{K} &= \widetilde{h}^{ij} \widetilde{K}_{ij}\\
    \Bar{K}_{ij} &= \frac{1}{2} \partial_r h_{ij} & \Bar{K} &= h^{ij} \Bar{K}_{ij} & \widetilde{\Bar{K}}_{ij} &= \frac{1}{2} \partial_\rho \widetilde{h}_{ij} & \widetilde{\Bar{K}} &= \widetilde{h}^{ij} \widetilde{\Bar{K}}_{ij}
\end{align}

The extrinsic curvature tensors transform under the coordinate transformation as
\begin{equation}
    \begin{split}
        K_{i j} &= e^{-\zeta} \widetilde{K}_{i j} +\mathcal{O}(\rho)~~~~~~   K = e^{-\zeta} \widetilde{K} +\mathcal{O}(\rho)\\
        \bar{K}_{i j}&=e^{\zeta } \widetilde{\bar{K}}_{i j}-e^{\zeta }\bigg[\tau  \left\{\xi _i \xi _j-\widetilde{\nabla _i}\xi _j+\frac{1}{2} \left(\xi _j \widetilde{\omega }_i+\xi _i \widetilde{\omega }_j\right) \right\}-\tau ^2 \bigg(-\frac{1}{2} \xi _a \xi _b \widetilde{h}^{a b} \widetilde{K}_{i j}+\xi _j \xi _k \widetilde{K}_i^k\\&+\xi _i \xi _k \widetilde{K}_j^k\bigg)\bigg] + \mathcal{O}(\rho) \\
        \bar{K}&=e^{\zeta } \widetilde{\bar{K}}-e^{\zeta }\left[\tau  \widetilde{h}^{i j}\left\{\xi _i \xi _j-\widetilde{\nabla _i}\xi _j+\xi _j \widetilde{\omega }_i 
        \right\}-\tau ^2 \left(2 \xi _i \xi _j \widetilde{K}^{i j}-\frac{1}{2} \widetilde{K} \xi _a \xi _b \widetilde{h}^{a b}\right)\right] + \mathcal{O}(\rho)
    \end{split}
\end{equation}
where the $\widetilde{\nabla}_i$s are covariant derivatives with respect to the induced metric $\widetilde{h}_{ij}$ in the $(\rho,\tau,y^a)$ coordinate system.

Some more structures that would be useful for calculating the change in entropy density and entropy current structures are given below.
\begin{equation}
\begin{split}
    \partial _v\bar{K}&=\partial _{\tau }\widetilde{\Bar{K}}-\left(\xi ^2-\widetilde{\nabla }\cdot \xi +\widetilde{\omega }\cdot \xi \right)\\
    &+\tau  \left(6 \xi \cdot \widetilde{K}\cdot \xi  -2 \xi _i\widetilde{\nabla }_j\widetilde{K}^{i j}-2 \widetilde{K}^{i j}\widetilde{\nabla }_i\xi _j+\xi \cdot \widetilde{\nabla }\widetilde{K}-\xi \cdot \partial _{\tau }\widetilde{\omega }+2 \widetilde{\omega }\cdot \widetilde{K}\cdot \xi -\xi ^2 \widetilde{K}\right) \\
    &+\tau ^2 \left(2\xi \cdot \partial _{\tau }\widetilde{K}\cdot \xi  - \frac{1}{2}\xi ^2 \partial _{\tau } \widetilde{K}\right) + \mathcal{O}(\rho)\\
    K \bar{K}&=\widetilde{K} \widetilde{\bar{K}}+ \tau \widetilde{K}\left(-\xi ^2+ \widetilde{\nabla }\cdot \xi -\widetilde{\omega }\cdot \xi \right) + \mathcal{O}(\rho) \\
    \nabla \cdot \omega &=\widetilde{\nabla }\cdot \widetilde{\omega }+2\widetilde{ \nabla }\cdot \xi \\
    &+\tau \left(-2\xi _i\widetilde{\nabla }_j\widetilde{K}^{i j}-2\widetilde{K}^{i j}\widetilde{\nabla }_j\xi _i+2 \xi \cdot \widetilde{K}\cdot \widetilde{\omega }-\widetilde{K} \widetilde{\omega }\cdot \xi +6 \xi \cdot \widetilde{K}\cdot \xi -2 \xi ^2 \widetilde{K}-\xi \cdot \partial _{\tau }\widetilde{\omega }\right)\\
    &+2 \tau ^2 \xi _i \xi _j \partial_\tau \widetilde{K}^{i j} \\
     \partial _{\tau }(\widetilde{\nabla} \cdot \xi )&=-2 \widetilde{K}^{i j} \widetilde{\nabla} _i\xi _j-2 \xi _i \widetilde{\nabla} _j\widetilde{K}^{i j}+\xi \cdot \widetilde{\nabla} \widetilde{K}  \\
     \omega ^2&=\widetilde{\omega }^2+4 \xi _a \left(\widetilde{h}^{a b} \left(\widetilde{\omega }_b+\xi _b\right)\right)-4 \tau  \xi _a \left(\widetilde{K}^{a b} \left(\widetilde{\omega }_b+2 \xi _b\right)\right) \\
     K^{i j} \bar{K}_{i j}&=\widetilde{K}^{i j} \widetilde{\bar{K}}_{i j}-\tau  \left(\xi _a \widetilde{\omega }_b \widetilde{K}^{a b}+\xi _a \xi _b \widetilde{K}^{a b}-\widetilde{K}^{a b}\widetilde{\nabla }_a\xi _b\right)
\end{split}
\end{equation}

The required Ricci and Riemann tensor components of the full metrics are expressed as
\begin{equation}\label{eq:ricciriemann}
    \begin{split}
        R_{r v}&=-\chi  - \frac{\omega ^2}{2} - \partial _v\bar{K} - K^{i j} \bar{K}_{i j}+ \frac{1}{2} \nabla \cdot \omega \\
         \widetilde{R}_{\rho  \tau }&=-\widetilde{\chi } - \frac{\widetilde{\omega }^2}{2} - \partial _{\tau }\widetilde{\bar{K}}- \widetilde{K}^{i j} \widetilde{\bar{K}}_{i j}+\frac{1}{2} \widetilde{\nabla }\cdot \widetilde{\omega } \\
         R_{rvrv} &= \chi + \frac{1}{4} \omega^2 \\
         \widetilde{R}_{\rho\tau\rho\tau} &= \widetilde{\chi} + \frac{1}{4} \widetilde{\omega}^2 \\
         R &= \mathcal{R} -2 \chi -\frac{3 \omega ^2}{2}-4 \partial_v \bar{K}+2\nabla \cdot \omega -2 K^{i j} \bar{K}_{i j}-2 K \bar{K} \\
         \widetilde{R} &= \widetilde{\mathcal{R}} -2 \widetilde{\chi} -\frac{3 \widetilde{\omega} ^2}{2}-4 \partial_\tau  \widetilde{\bar{K}}+2\widetilde{\nabla} \cdot \widetilde{\omega} -2 \widetilde{K}^{i j} \widetilde{\bar{K}}_{i j}-2 \widetilde{K} \widetilde{\bar{K}}
    \end{split}
\end{equation}

The Ricci scalar corresponding to the induced metric $\widetilde{h}_{ij}$ transforms on the horizon as
\begin{equation}
    \begin{split}
        \mathcal{R}-\widetilde{\mathcal{R}}= -2\tau  \left( -\xi \cdot \widetilde{K}\cdot \xi   + \xi ^2 \widetilde{K} + 2 \xi _i\widetilde{\nabla _j}\widetilde{K}^{i j}+ \widetilde{K}^{i j}\widetilde{\nabla _j}\xi _i - \widetilde{K} \widetilde{\nabla }\cdot \xi -2\xi \cdot \widetilde{\nabla} \widetilde{K} \right) \\ - 2 \tau ^2 \left(\xi ^2 \partial_\tau \widetilde{K}-\xi \cdot \partial_\tau \widetilde{K}\cdot \xi \right)
    \end{split}
\end{equation}

\bibliographystyle{JHEP}
\bibliography{ref}

\end{document}